\documentclass[final,5p,times,twocolumn,authoryear]{elsarticle}

\usepackage{amssymb}
\usepackage{lipsum}

\usepackage[nogroupskip, nopostdot]{glossaries}
\usepackage{glossaries-extra}
\usepackage{fdsymbol}

\usepackage{soul}

\usepackage{ulem}

\makeglossaries
\setabbreviationstyle{long-short}
\newabbreviation{3d}{3D}{Three-Dimensional}
\newabbreviation{3gpp}{3GPP}{Third Generation Partnership Project}
\newabbreviation{4g}{4G}{Fourth Generation}
\newabbreviation{5g}{5G}{Fifth Generation}
\newabbreviation{5g-cpe}{5G-CPE}{5G Customer Premise Equipment}
\newabbreviation{6g}{6G}{Sixth Generation}
\newabbreviation{ai}{AI}{Artificial Intelligence}
\newabbreviation{afar}{AFAR}{Forward Area Based Routing-algorithm}
\newabbreviation{afw}{AFW}{Autonomous Flight Wireless}
\newabbreviation{ao}{AO}{Alternating Optimization}
\newabbreviation{bcot}{BCoT}{Blockchain of Things}
\newabbreviation{cdf}{CDFs}{Cumulative Distribution Functions }
\newabbreviation{cr}{CR}{Cognitive Radio}
\newabbreviation{cmp}{CMP}{Coordinated Multiple Points}
\newabbreviation{csi}{CSI}{Channel State Information}
\newabbreviation{dc}{DC}{Difference-of-Concave}
\newabbreviation{dlt}{DLT}{Distributed Ledger Technology}
\newabbreviation{dtn}{DTN}{Delay Tolerant Networks}
\newabbreviation{dqn}{DQN}{Deep Q-Networks}
\newabbreviation{dsm}{DSM}{Dynamic Spectrum Management}
\newabbreviation{gis}{GIS}{Geographic Information Systems}
\newabbreviation{g2u}{G2U}{Ground-to-UAV}
\newabbreviation{gt}{GTs}{Ground Terminals}
\newabbreviation{gps}{GPS}{Global Positioning System}
\newabbreviation{hppp}{HPPP}{Homogeneous Poisson Point Process}
\newabbreviation{iaco}{IACO}{Improved Ant Colony Optimization}
\newabbreviation{ip}{IP}{Intercept Probability}
\newabbreviation{ioe}{IoE}{Internet of Everything}
\newabbreviation{ioit}{IoIT}{Internet of Intelligent Things}
\newabbreviation{iot}{IoT}{Internet of Things}
\newabbreviation{kkk}{KKK}{Karush-Kuhn-Tucker}
\newabbreviation{los}{LOS}{Line of Sight}
\newabbreviation{lte}{LTE}{Long Term Evolution}
\newabbreviation{m2m}{M2M}{Machine to Machine}
\newabbreviation{mbb}{MBB}{Mobile Broadband}
\newabbreviation{mec}{MEC}{Mobile Edge Computing}
\newabbreviation{misome}{MISOME}{Multiple-Input-Single-Output Multiple-Eavesdropper}
\newabbreviation{mimo}{MIMO}{Multi-Input Multi-Output}
\newabbreviation{ml}{ML}{Machine Learning}
\newabbreviation{nfz}{NFZs}{No-Fly Zones}
\newabbreviation{ndn}{NDN}{Never Die Networks}
\newabbreviation{noma}{NOMA}{Non-Orthogonal Multiple Access}
\newabbreviation{ofdm}{OFDM}{Orthogonal Frequency-Division Multiplexing}
\newabbreviation{pow}{PoW}{Proof-of-Work}
\newabbreviation{qos}{QoS}{Quality of Service}
\newabbreviation{ran}{RAN}{Radio Access Network}
\newabbreviation{rf}{RF}{Radio Frequency}
\newabbreviation{rmicn}{RMICN}{Router-movable Information-centric Networking}
\newabbreviation{scp}{SCP}{Secure Connection Probability}
\newabbreviation{seem}{SEEM}{Secrecy Energy Efficiency Maximization}
\newabbreviation{smf}{SMF}{Session Management Function}
\newabbreviation{sdn}{SDN}{Software-Defined Networking}
\newabbreviation{slam}{SLAM}{Simultaneous Localization and Mapping}
\newabbreviation{snr}{SNR}{Signal-to-Noise Ratio}
\newabbreviation{sca}{SCA}{Successive Convex Approximation}
\newabbreviation{swipt}{SWIPT}{Simultaneous Wireless Information and Power Transfer}
\newabbreviation{thz}{THz}{Terahertz}
\newabbreviation{tdma}{TDMA}{Time Division Multiple Access}
\newabbreviation{uav}{UAV}{Unmanned Aerial Vehicle}
\newabbreviation{vr}{VR}{Virtual Reality}
\newabbreviation{u2g}{U2G}{UAV-to-Ground}
\newabbreviation{upf}{UPF}{User Plane Function}
\newabbreviation{wifi}{WiFi}{Wireless Fidelity}
\newabbreviation{wsn}{WSN}{wireless sensor network}
\newabbreviation{wcsr}{WCSR}{Worst-Case Secrecy Rate}
\newabbreviation{wimax}{WiMAX}{Worldwide interoperability for Microwave Access Network}
\setglossarystyle{tree}

\journal{Computers \& Security}

\begin{document}

\begin{frontmatter}



\title{Blockchain-Based Security Architecture for Unmanned Aerial Vehicles in B5G/6G Services and Beyond: A Comprehensive Approach}

\author[label1]{Senthil Kumar Jagatheesaperumal}
\affiliation[label1]{organization={Department of Electronics and Communication Engineering, Mepco Schlenk Engineering College},
            city={Sivakasi},
            state={Tamil Nadu},
            country={India}}


\author[label2]{Mohamed Rahouti}
\affiliation[label2]{organization={Department of Computer and Information Sciences, Fordham University},
            city={New York},
            postcode={10023}, 
            state={NY},
            country={USA}} 

\author[label3]{Kaiqi Xiong}
\affiliation[label3]{organization={Cyber Florida, University of South Florida},
            city={Tampa},
            postcode={33620}, 
            state={FL},
            country={USA}}

\author[label4]{Abdellah Chehri}
\affiliation[label4]{organization={Royal Military College of Canada},
            city={Kingston},
            postcode={K7K 7B4}, 
            state={Ontario},
            country={Canada}}

\author[label5]{Nasir Ghani}
\affiliation[label5]{organization={Electrical Engineering Department, University of South Florida},
            city={Tampa},
            postcode={33620}, 
            state={FL},
            country={USA}}

\author[label2]{Jan Bieniek}

\begin{abstract}
Unmanned Aerial Vehicles (UAVs), previously favored by enthusiasts, have evolved into indispensable tools for effectively managing disasters and responding to emergencies. For example, one of their most critical applications is to provide seamless wireless communication services in remote rural areas. Thus, it is substantial to identify and consider the different security challenges in the research and development associated with advanced UAV-based B5G/6G architectures. Following this requirement, the present study thoroughly examines the security considerations about UAVs in relation to the architectural framework of the 5G/6G system, the technologies that facilitate its operation, and the concerns surrounding privacy. It exhibits security integration at all the protocol stack layers and analyzes the existing mechanisms to secure UAV-based B5G/6G communications and its energy and power optimization factors. Last, this article also summarizes modern technological trends for establishing security and protecting UAV-based systems, along with the open challenges and strategies for future research work.
\end{abstract}


\end{frontmatter}

\section{Introduction} \label{sec:intro}

Unmanned aerial vehicles (UAVs) or drones have attracted rising attention towards wireless communications applications owing to their numerous advantages, such as easy mobility control and establishment of required communication links at low cost. They can also be used as an aerial base station to enhance wireless capacity, coverage, and energy efficiency. UAVs can be like small aircraft balloons or drones; they can be remotely controlled or pre-programmed. UAVs have many applications in military, surveillance, search and rescue localization, and telecommunications. Technical challenges exist in providing an analytical framework and efficient algorithms that can be used to analyze and optimize the design of drone-based communication systems.

Under Project SkyBender, Google experimented with drones to deliver high-speed 5G Internet services ~\cite{zhou2018air}. The millimeter-wave technology has the potential to transmit gigabytes of data every second with speeds up to 40 times faster than 4G internet. However, the millimeter-wave signals are absorbed by atmospheric gases, while rain and humidity can also reduce signal strength. Millimeter waves also travel via line of sight, which means objects, including trees and buildings, can blog them.

A blockchain system enhances the security and dependability of the underlying infrastructure without relying on human trust \cite{rahouti2018bitcoin}. Blockchain networks are built in such a way that it is assumed any individual node could attack them at any time. Consensus protocols like proof-of-work (PoW) ensure that even if the attack happens, the network completes its functions as intended, regardless of human cheating or intervention. The blockchain allows one to store data and secure it using various cryptographic properties such as digital signatures and hashing as soon as data enters a block in a blockchain network. If intruders try to tamper with the blockchain database, consensus would recognize and shut down the attempt. The consensus prevents data tampering and network attacks.

Blockchains are made up of nodes, which can be within one institution, or multiple institutions all over the world on the computer of any citizen that wants to participate in any decision to be made. They form a democratic system, so if anyone node is compromised and malicious actions are taken, the other nodes would automatically recognize the problem and simply not execute the inappropriate activity. The blockchain has incredible security features, so everyone uses it to store essential data correctly. One of the striking features of blockchains is their scalability features, in which transactions per second measure blockchains with nodes all over the world run by independent citizens.

Using UAVs as a service (e.g., Drone-as-a-Service) is a revolutionary development for the proposition, similar to the Infrastructure-as-a-Service model, UAV-as-a-Service would be directly structured to perform UAV tasks and responsibilities \cite{nguyen2023advanced}. With the inclusion of blockchain, businesses can benefit from increased continuity by providing that service. Applicable technologies such as 5G and 6G allow for intercommunication between multiple parties, enabling users of UAV technology to benefit from cross-functional use cases that serve as a service platform. With UAVs on the horizon, there is potential for market capital in opportunistic delivery to civilians without needing a driver, thereby serving society by completing tasks and assignments.

Furthermore, adopting blockchain technology into UAV environments enables a "zero-trust policy," ensuring that all parties can see transparency in the service and data provided. With both UAVs and the Internet of Things (IoT) architectures still in their adolescence, there is no plausible way to justify keeping private data away from its viewers. The public blockchain is known to be a public structure, which further emphasizes the need for a zero-trust policy. Allowing for zero trust would build trust in the overall ecosystem nurtured by all parties. Specifically, zero-trust security frameworks can authenticate, authorize, and validate all users, thereby sourcing the responsibility for all actions and outcomes \cite{yang2023toward}. Congruent methods to this approach would also be valid examples of potential security solutions for the overall security concept in UAV infrastructure.

\subsection{Contributions of this work}

In this article, a detailed investigation and analysis of the usage of blockchain security solutions have been carried out for UAV communication. The motivation behind the compilation of the recent research work towards UAVs in 5G/6G environments and applications is to encourage researchers to develop better connectivity systems even in remote areas. The review provides awareness about the state-of-the-art modern communication services and associated solutions for UAV communications. It intends to offer state-of-the-art blockchain-enabled security frameworks with the support of modern enabling technologies, relevant approaches, and future perspectives. The major contributions of this article are summarized as follows.

\begin{itemize}
\item To provide a detailed taxonomy of the evolution of 5G/6G services and their applications to UAVs.
\item To provide a deep discussion and insight into using UAVs for communication services and domains.
\item To provide a deep discussion and insight on the use of security concerns of UAV communication services. 
\item To provide a deep discussion and insight into modern technological trends supporting energy and power-optimized solutions.
\item To provide an exhaustive summary of the state-of-the-art algorithms for security enhancement in UAV communication.
\item To provide the blockchain-enabled security solution for UAV communication.
\item To provide a few case studies on enhancing security aspects in blockchain-based solutions to understand the subject of discussions better.
\item To provide the state-of-the-art research challenges and the comparative discussion on existing articles.
\end{itemize}


\subsection{Organization of the article}
This review article is organized as follows. Section \ref{sec:survey} summarizes the related survey towards the chosen domain. Section \ref{sec:taxanomy} discusses the evolution of B5G/6G Communication services from 5G services in the market. It summarizes the application of UAVs in different domains and their unique characteristics towards communication services. Section \ref{sec:security} discusses the security concerns of UAV-based architectures and the appropriate measures to ensure privacy and security in communication. Section \ref{sec:security-support} summarizes the energy and power optimization requirement for security enforcement in UAV-based Architectures. In Section \ref{sec:blockchain}, Blockchain-based security architecture for UAV in B5G/6G is proposed with detailed architecture and summary. Specific open issues, challenges, and applications in the security aspects of UAV architecture are discussed in Section \ref{sec:issues} along with the key findings and summary of the article towards establishing the best and secure means of UAV-based architectures. It also encapsulates the recent trends and future scope of research left as open-end challenges for researchers. Finally, Section \ref{sec:conclusion} concludes this work.


\section{Summary of Related Surveys from Literature} \label{sec:survey}

This section deals with the compilation of the contributions of various researchers toward the growth and usage of UAVs for various application domains. This section also summarizes the 5G services and architectures from the literature. Furthermore, blockchain-related surveys towards security enforcement for different applications are discussed from the surveys' outcomes, and the potential research gaps are pointed out for elaboration in the rest of this survey article. The few most promising survey articles related to the scope of this article are compiled and presented with their applications next.

Fotouhi~\textit{et al.}~\cite{fotouhi_survey_2019} conducted a systematic literature review on UAV cellular communications to better integrate UAVs in cellular networks. The authors also reviewed types of consumer UAVs, cyber-physical security, ongoing prototyping, challenges, and opportunities for assisting cellular communications. A recent survey of the Physical layer security techniques on passive wireless eavesdropping and its applications is available in~\cite{hamamreh_classifications_2019}. In this review, the authors also discuss the expandable framework for future research directions and the support of IoT and 5G technologies for future wireless systems.

Fight against active and passive eavesdropping in UAVs and potential research challenges described in this section has been proposed in~\cite{sun_physical_2019}, where Sun et al. also suggest using trajectory design and cooperative UAVs Physical Layer Security in UAVs can be ensured with improved spectral efficiency. The survey by Wang~\textit{et al.}~\cite{wang_uav-involved_2019} covers the secure wireless communications involved in the physical-Layer of UAVs that focus on security aspects considering various roles of a UAV with optimization at the degree of freedom level.

The survey by Wang~\textit{et al.}~\cite{wang_physical-layer_2019} covers the Physical Layer Security aspects of 5G wireless communication for IoT devices along with their application scenarios, security threats, and communication technologies in 5G towards IoT, and management of physical-layer threats. The survey conducted in~\cite{wu_safeguarding_2019} includes the mechanisms for tackling the new issues that arise in the physical-layer security of wireless communication networks with validated numerical results and future direction for research.

Case studies on UAVs in aerial mode and flying base station modes with the demonstration of physical layer security mechanisms providing superior performance gains are summarized in~\cite{li_secure_2019}. In the studies, securing communications is ensured over 5G networks, and also emerging future research directions are highlighted.

\begin{table*}[h!]
\caption{Existing surveys on most relevant secure communication architectures for UAVs.} 
\label{tab:surveys}
\centering
\begin{tabular}{|p{2.4cm}|p{1cm}|p{2.0cm}|p{6.5cm}|p{4.0cm}|}
\hline
\textbf{Related \newline works} 
& \textbf{Year} 
& \textbf{Topic} 
& \textbf{Key contributions}
& \textbf{Research gaps}\\ \hline
\hline
\cite{chen2020review}  & 2020 & Swarm communication  &  Surveys the UAV swarm routing protocols and communication architectures.  &  No direct focus on security.   \\ \hline
\cite{hentati2020comprehensive}  & 2020 & Communication networks  &  Surveys the UAV communication protocols, architectures, and their applications. & No discussion on secure deployments.   \\ \hline
\cite{gupta2021fusion} & 2021 &  Drone networking  &  Surveys secure UAV communication with the fusion of AI and Blockchain. & No in-depth coverage on security among a wide range of applications.  \\ \hline
\cite{ullah2020cognition}  & 2020  & Cognition aspects  &  Studies the cognition capabilities of the UAV-based communication. & No wide focus on security aspects.  \\ \hline
\cite{shakoor2019role}  & 2019 & Public safety &  Summarizes the impact of UAVs in public safety communications. & Addresses safety only from the perspective of energy efficiency.   \\ \hline
\cite{aggarwal2020blockchain} & 2020 & 6G networks  &  Surveys on 6G use cases on the taxonomy of the UAV communication. & No stringent efforts on addressing security challenges. \\ \hline
\cite{mccoy2019software} & 2019 & SDN  &  Surveys on SDN-based solutions for addressing vulnerabilities in the UAV network. & No direct focus on the range of security measures.  \\ \hline
\cite{nawaz2021uav}  & 2021 & Communication networks  &  Surveys on the issues in UAV communication networks. & Security aspects are least considered. \\ \hline
\cite{noor2020review}  & 2020 & Ad-hoc networks  &  Reviews on the key enabling communications technologies of flying ad-hoc networks. & No considerable focus on security.  \\ \hline
\cite{qadir2021addressing}  & 2021 & Path planning   &  Reviews on the UAV communication network employed for disaster management in smart cities. & No key focus on security from the UAV perspective.   \\ \hline
\cite{al2020uavs} & 2020 & SDN IoT  &  Surveys on the applications of the UAV base station enabled with SDN support. & Only on-demand security concerns are highlighted.\\ \hline
\cite{zhi2020security} & 2020 & Security and privacy  & Surveys on the UAV safety aspects considering communication, sensors, and multi-UAVs. & No potential impact on the security of multi-UAVs.   \\ \hline
\cite{vashisht2020mac} & 2020 & MAC protocols &  Reviews on the support of MAC protocols for UAV-based communication systems. & Security aspects are not highlighted for the prescribed ecosystem.   \\ \hline
\cite{chriki2019fanet} & 2019 & Ad-Hoc network  &  Surveys on the existing communications architectures for the flying ad-hoc networks. & Very sparse focus on the security challenges.   \\ \hline
\cite{wheeb2022topology}  & 2022 & Routing Protocols and Mobility  &  Reviews on the mobility models and routing protocol for flying ad hoc networks. & Lacks discussion on the security aspects of the protocols.   \\ \hline
\cite{rugo2022security}  & 2022 & UAV network security  &  Reviews on the security aspect of UAV networks considering a diversified range of parameters. & Security aspects not focused on Blockchain.  \\ \hline
\cite{duong2022uav}  & 2022 & Caching in 6G networks & Surveys on the analyses of cache models for the UAV communication through 6G services. & No considerable focus on security aspects.     \\ \hline
\cite{siddiqi2022analysis}  & 2022 & Security analysis & Analyses security and privacy concerns of UAVs subjected to the prescribed regulation. & Least focus on communication security.   \\ \hline
Our Paper & 2023 & Blockchain for securing UAV communication & A comprehensive survey on Blockchain and 5G/6G services integration for UAVs; The roles of Blockchain in various key 5G/6G-enabled UAV systems; The use of Blockchain in dependable 5G applications, including edge devices and their potential to meet 6G requirements. &  - \\ \hline
\end{tabular}
\end{table*}
\begin{figure*}[!t]
\centering
\centerline{\includegraphics[height=12cm, width=16cm]{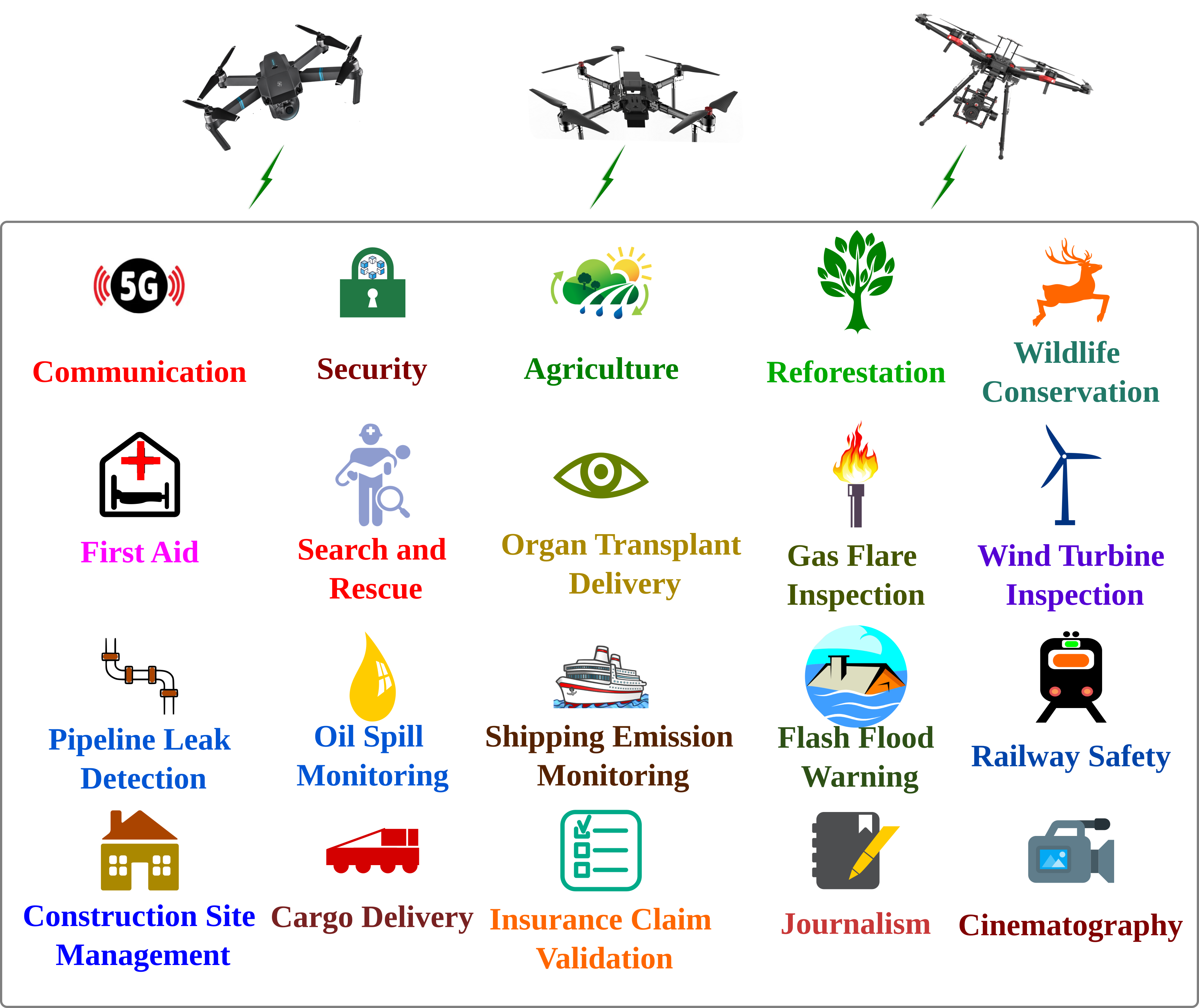}}
\caption{Taxonomy of Commercial and General use cases of UAVs.}
\label{fig:usecases}
\end{figure*}


\section{A Detailed Taxonomy on the Evolution of B5G/6G Communication Services}
\label{sec:taxanomy}

This section provides a comprehensive overview of the state-of-the-art existing works related to B5G/6G-based communication services and their evolution. Communication with UAVs can be established through WiFi networks or private wireless networks. In the case of connected UAVs, they use cellphone network services such as 5G/B5G/6G to communicate with the controller. Generally, connected UAVs have six modules in these systems. 
\begin{itemize}
    \item \textit{Communication System:} The UAVs need to communicate with the network and receive the control command from the network. They also are engaged in uploading pictures and videos to the network. Such UAVs for establishing a connection with the network must have a sim card with 5G/B5G/6G support. Particularly for establishing B5G/6G services, the UAV must be deployed with B5G/6G customer premise equipment (CPE), which acts as a gateway for integrating UAVs with the environment and the controller. 
    
    \item \textit{Flight control system:} For controlling the position of the UAVs navigating in the air, we need to have a flight control system. They must possess higher reliability and higher stability with lightweight. They must be intelligent and efficient enough to control the positions of UAVs. They usually receive the command from the ground controller through the network. Based on the commands, the UAVs avoid obstacles and maneuver through the desired path.
    
    \item \textit{Navigation System:} The navigation system helps plan the paths for the UAVs. It should possess low latencies, large bandwidth, and provision over-the-horizon remote control. It should also ensure to provide autonomous navigation through appropriate path-planning algorithms. High-precision positioning and cluster flight are other vital features of the navigation system that could assist UAVs in finding the best paths and routes.
    
    \item \textit{Flight Management System:} This subsystem includes authentication and security encryptions, so we cannot allow the UAVs to be controlled by unauthorized persons. 
    
    \item \textit{Task Load System}: Most of the UAVs that would be employed for navigation can carry some lightweight objects such as cameras and food items. The objects should be lightweight and miniature airborne equipment. The task load systems with a diversified range of airborne equipment will be in much demand for commercial applications. Further, the real-time network transmission should also be sustained with the load in the UAVs.
    
    \item \textit{Airborne Computers:} Sensing and other intelligent observations from the UAV environment during its navigation could be programmed in an airborne computer. The data accumulated from the environment could be used for the second level of application development with the support of the programmed airborne computer.
\end{itemize}
\subsection{Role of UAVs in General and Communication Services } \label{sec:roleofuav}

The modules and protocols in these networks are of the most significant importance to enable a reliable communication network for UAVs. Several methods have been introduced in the past by both research and industry communities. However, the vast majority of these research efforts only consider a few critical communication factors, including resource management, antenna design, and communication network architecture. Thus, this section will review and discuss the communication modules related to UAVs and their associated networking paradigms. Further, a thorough comparison will be elaborated on the mechanisms and algorithms used in 5G-enabled UAV communications.

\subsubsection{5G Architecture}
The Radio Access Network (RAN) belonging to 5G network services, popularly known as new radio, has been standardized to allow tight internetworking with the existing LTE services. 5G network architectures enabling tight internetworking among LTE and new radio services are known as non-standard architectures. These allow a smooth and relatively simple evolution toward a complete end-to-end 5G system. They enable the reuse of the existing LTE-based base stations and 4G core networks by a simple software upgrade among the new radio base stations. The combination of a standalone new radio base station and a 5G core network is known as a 5G system. 

\begin{figure}[!t]
\centering
\centerline{\includegraphics[width=0.5\textwidth]{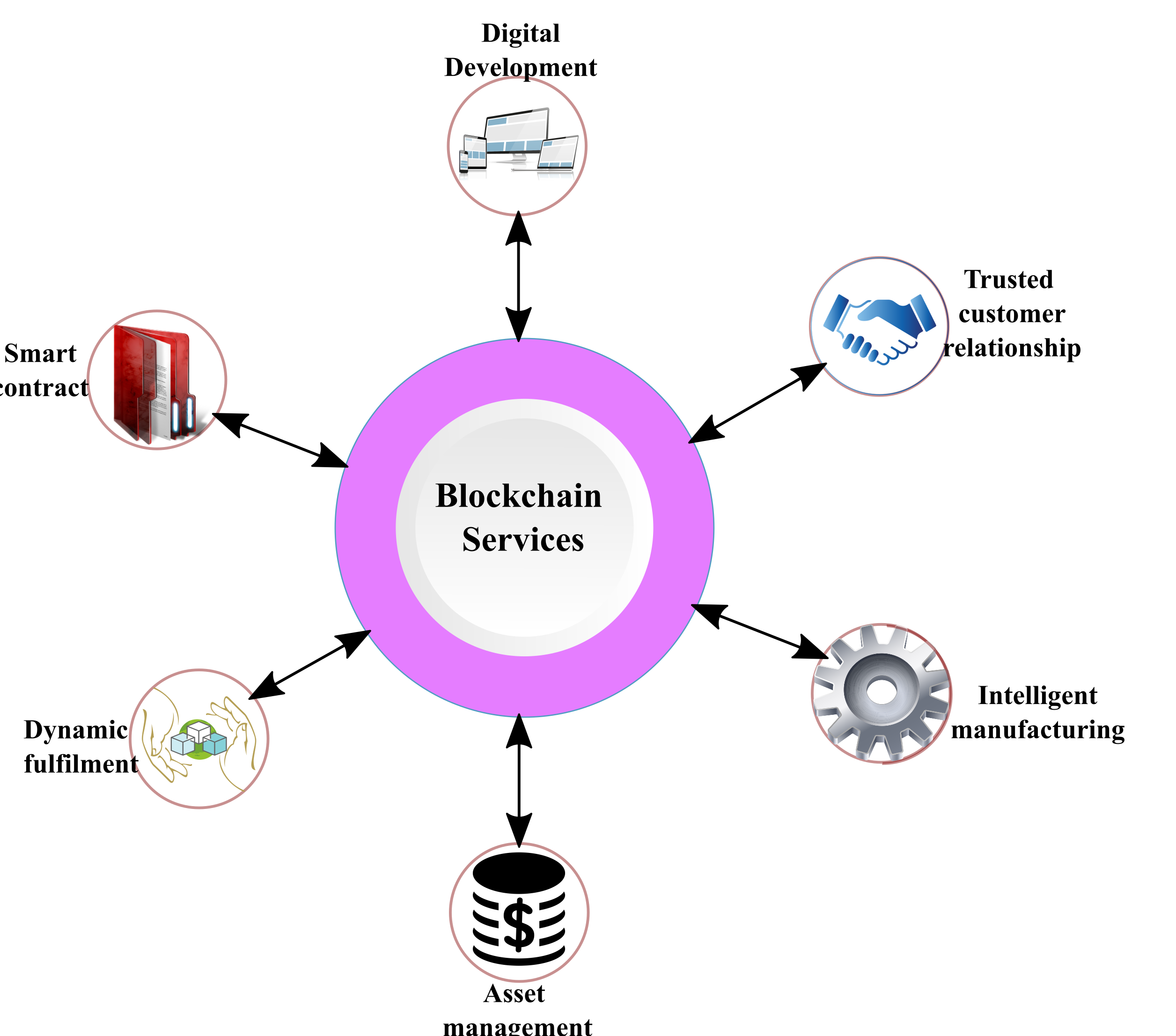}}
\caption{Trustworthy blockchain services.}
\label{fig:5Gusage2}
\end{figure}

Each use case category of 5G includes many individual applications, as shown in Fig.~\ref{fig:5Gusage2}. The existing applications included voice service, video streaming, Internet access, social media, and instant message applications. Many emerging applications include automatic vehicles, remote home automation, smart cities, and smart wearables. Many of these applications require components from multiple use case categories. For instance, virtual reality (VR) requires low latency to provide high mobile broadband (MBB) and responsiveness throughputs for rapid content transfer. 3GPP ensures that 5G services have sufficient flexibility and capability to address the diverse requirements of this broad range of applications. Also, 3GPP has specified reference-based and service-based architecture for the 5G systems. 

\begin{figure*}[!t]
\centering
\centerline{\includegraphics[height=12cm, width=16cm]{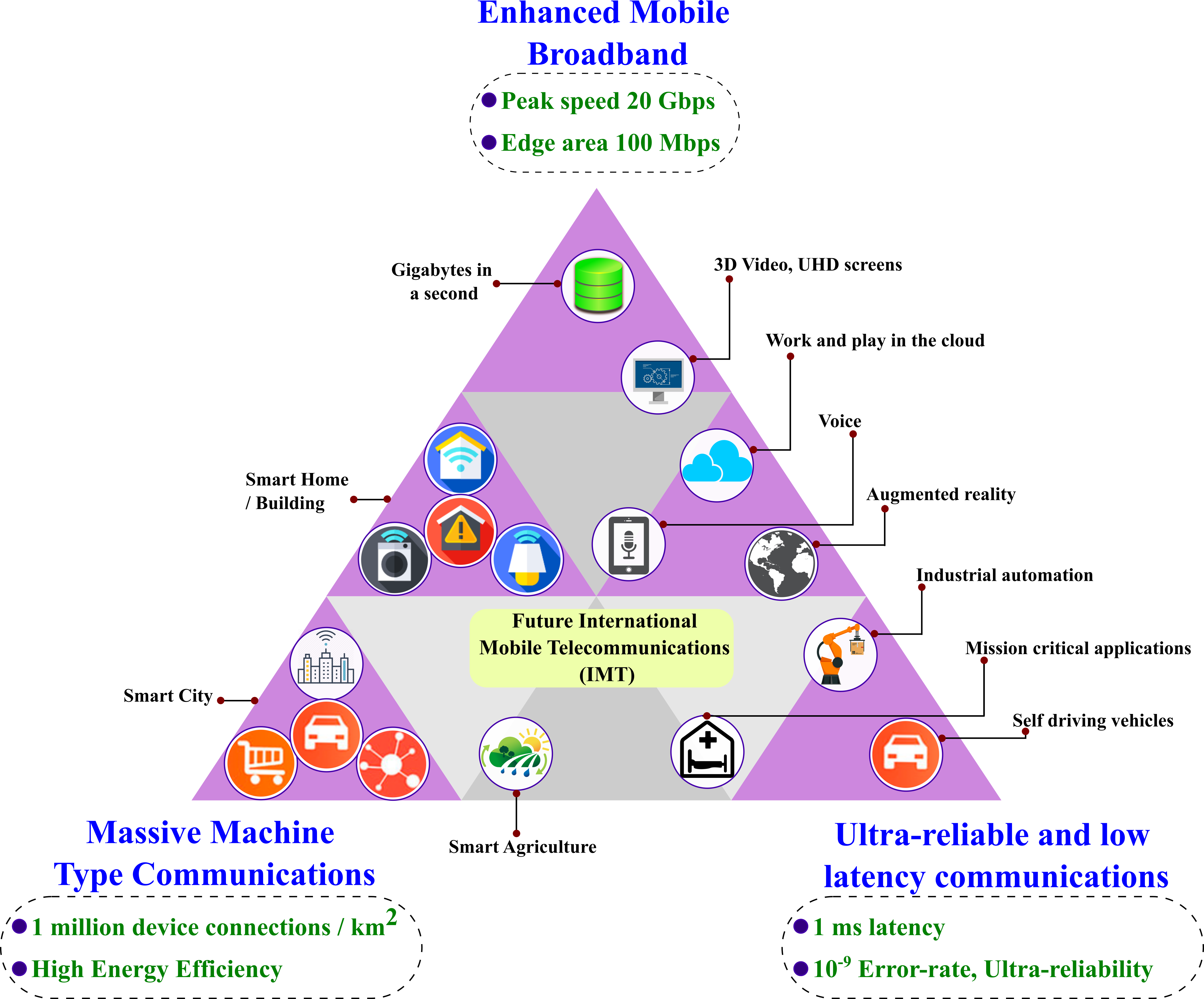}}
\caption{The 5G technology usage scenarios.}
\label{fig:5Gusage}
\end{figure*}

The reference-based architecture is based on network elements that use point-to-point interfaces to interconnect those network elements. Signaling procedures are specified for each point-to-point interface. It leads to repetition with the specifications if the same signaling procedure is used across multiple interfaces. The service-based architecture replaces the network elements set with the network functions. Each network function can provide services to other functions by acting as a service provider. Here, the point-to-point interfaces are replaced by a common bus that interconnects all the network functions. One of the crucial characteristics of a 5G system is the separation of the user plane and control plane functions. The session management function (SMF) is used for IP address allocation, and the user plane function (UPF) is meant for packet forwarding in the 5G system architecture. This supports the independent scaling of these two functions because of the separation of the user and control planes. This feature allows the operators to add more user plane capability without more control plane capability. Furthermore, the user plane functions could be distributed, and the control plane functions could be centralized. 
\begin{figure*}[!t]
\centering
\centerline{\includegraphics[height=7cm, width=15cm]{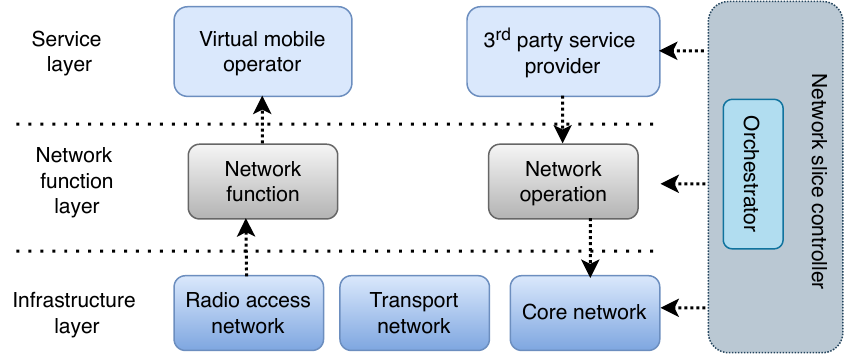}}
\caption{The architecture of 5G technology (network slicing).}
\label{fig:5G-architecture}
\end{figure*}

\subsubsection{Networking Technologies in UAV Communication Systems}
Notable research works have been conducted in the past with a focus on the various aspects of technologies-related communication networks of drones, enabling UAV technological enhancement with robust and resilient networks. Among these works, Rahman et al.~\cite{rahman2014enabling} notably considered worldwide interoperability for microwave access network (WiMAX) as a proper technology for examining different wireless-based communication protocols that are based on the SHERPA network standard criteria, including ZigBee, WiFi, and XBee. Furthermore, the authors in \cite{lee2016devising} deployed the Forward Area Based Routing-algorithm (AFAR) along with Geographic Information Systems (GIS) to design an adaptive solution for drone communication improvement.

Moreover, the authors in~\cite{sharma2020communication} proposed Router-movable Information-centric Networking (RMICN), a communication system to enable disjoint-pair communication among drones. Their solution leverages the relay nodes and the movement of flying routers' physical control to ensure efficiency and communication reliability. Further studies such as~\cite{akka2018mobile} addressed the path planning strategies problem in drones and proposed Improved Ant Colony Optimization (IACO), a path planning solution for a set of mobile robots.

As resource allocation is another vital aspect of communication technologies of UAVs, the authors in~\cite{yoshikawa2017resource} aimed to explore the optimal frequency band to provide communication efficiency and optimization while avoiding interference in the case of simultaneous communications. This work further enhances the usage of the primary band used by the individual drones in accordance with the total number of drones. Other works such as~\cite{fabra2017impact} also addressed the interference problem in drone communication. The authors in~\cite{fabra2017impact}, in particular, examined the high interference problem at the radio control and presented a comprehensive evaluation of the existing interference optimization techniques. Their results validated the incompatibility of WiFi with the radio control unit in the 2.4 GHz wireless band. This is mainly due to the high number of remote control devices that use this communication band.

Building upon this, the authors in~\cite{shrit2017new} developed a lightweight framework to facilitate drone synchronization using ad-hoc communication when forming a swarm network. Here, to operate a swarm network, the drones autonomously follow the lead of an individual leader drone, which a human must pilot.

Last, self-recovery network technologies have drawn remarkable attention to the UAV domain. For instance, Uchida et al.~\cite{uchida2014evaluation} proposed a self-recovery-based resilient communication solution comprising Delay Tolerant Networks (DTN), Never Die Networks (NDN), and Autonomous Flight Wireless (AFW) devices to provide reliable and resilient communication with wireless stations in isolated geographical regions.

\subsubsection{UAV-Assisted WSN and UAV-Assisted Vehicular Communication Systems}
The deployment of a Wireless sensor network (WSN) for UAV environments is challenging because of the distributed localization of sensors in a large geographical area (i.e., dense sensors). It has been demonstrated that the static-based integration of WSN is significantly less efficient, and it incurs progressive phases of damage or disaster~\cite{erdelj2017help}. To alleviate and manage potential disasters, recommendations are provided for the UAV and WSN based on three key classifications: pre-disaster preparedness, disaster assessment, and disaster response/response. A notable research work by Wu et al.~\cite{wu2017orsca} suggested incorporating mobile data in WSN by a UAV, where a regulated greedy-based routing mechanism was leveraged to address the problem of route selection and communication association. In~\cite{sharma2020communication}, the authors explored the possible means of communication technologies for UAVs through various ranges of network architectures, resource management platforms, and antennas. Further, the interplay between the IoT networks for advanced cellular communication for UAVs is focused on a diversified range of UAV use cases, from navigation to surveillance.

\subsection{B5G/6G-enabled UAV Communications}
Recent communication technologies have started exploiting the potential of using B5G/6G technology to enable efficient and secure communication among UAVs. This section highlights the key features of 6G technology, which can allow UAVs to operate at higher speeds, cover larger distances, and perform more complex tasks. It also discusses the challenges and opportunities associated with B5G/6G-enabled UAV communication, including spectrum allocation, network architecture, and security considerations.

\subsubsection{Towards B5G/6G with Connected Sky}

\begin{figure*}[!t]
\centering
\centerline{\includegraphics[height=12cm, width=10cm]{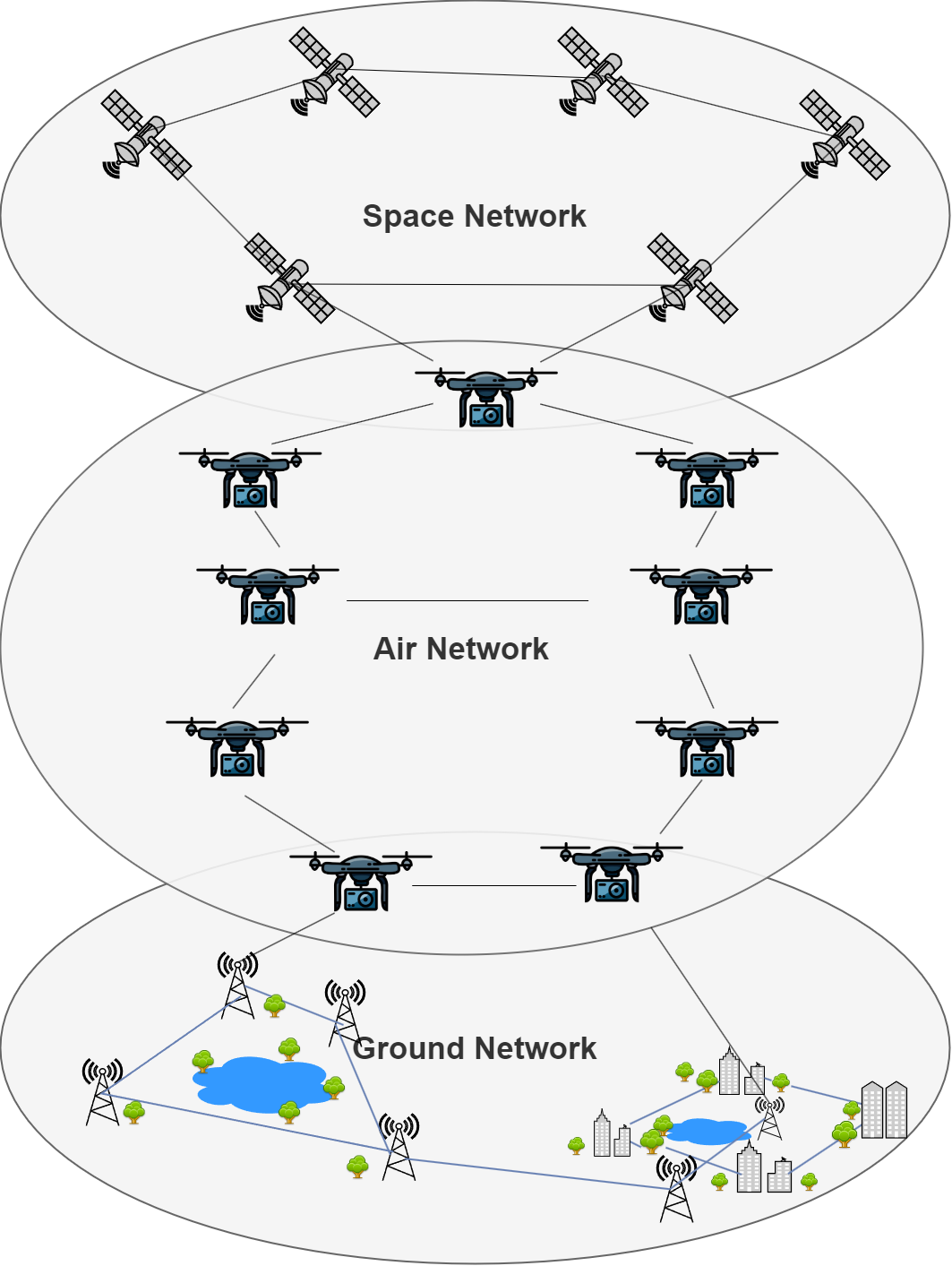}}
\caption{B5G/6G technology-supported UAV communications architecture.}
\label{fig:6G}
\end{figure*}

B5G/6G technology promises to bring many advanced features and provide faster data transfer rates, low latency, and improved network coverage. Future connected societies will be powered by this technology, enabling virtual and augmented reality, autonomous vehicles, and IoT. Secure communication is an essential aspect of B5G/6G technology~\cite{de2021survey}. It is expected that B5G/6G technology will provide a secure and reliable communication infrastructure for users who have increased concerns about privacy and security in the era of digital communication. It is also designed to address the challenges posed by current wireless communication technologies, including the high risk of cyber-attacks and data breaches~\cite{zhou2021fine}. This is achieved through advanced security protocols and encryption technologies that protect against unauthorized access and ensure data privacy in transit. The protocols are designed to detect and prevent malicious activities such as man-in-the-middle attacks, denial-of-service attacks, and other hacking attempts.

\begin{table*}[h!]
\centering
\caption{Summary on state-of-art usage of modern communication services for UAVs and their limitations.} 
\label{tab:CommunicationUAV}
\begin{tabular}{|p{2.2cm}|p{0.5cm}|p{1.5cm}|p{1.8cm}|p{4.2cm}|p{4.2cm}|}
\hline
\textbf{References}        & \textbf{Year} & \textbf{Single/Multi UAVs} & \textbf{Communication Network} & \textbf{Communication Strategy}                 & \textbf{Limitations}                         \\ \hline
Yu et al.~\cite{yu2021uav}     & 2021          & Multi   UAVs               & 5G, B5G               & Multi-hop   relay, multi-access edge computing & Limited   coverage area, high latency        \\ \hline
Shnaiwer et al.~\cite{shnaiwer2022multi}   & 2022          & Multi   UAVs               & 5G, 6G                & Multi-hop   relay, beamforming, edge computing & Limited   coverage area, high latency        \\ \hline
Pan et al.~\cite{pan2021uav} & 2021          & Single   UAV               & 5G                    & Single-hop   relay                             & Limited   coverage area, high latency        \\ \hline
Zhao et al.~\cite{zhao2019uav}    & 2019         & Single   UAV               & 5G                    & Direct   communication, beamforming            & Limited   coverage area, high latency        \\ \hline
He et al.~\cite{he2021multi}   & 2021         & Single   UAV               & 5G                    & Direct   communication, multi-hop relay        & Limited   coverage area, high latency        \\ \hline
Amponis et al.~\cite{amponis2022drones}    & 2022          & Multi   UAVs               & B5G                   & Multi-hop   relay, edge computing              & Limited   coverage area, high latency        \\ \hline
Qi et al.~\cite{qi2019uav}  & 2019          & Single   UAV               & 5G                    & Direct   communication, beamforming            & Limited   battery life                       \\ \hline
Alsamhi et al.~\cite{alsamhi2022computing}    & 2022         & Multi   UAVs               & 6G                    & Multi-hop   relay, edge computing              & Still   in development, limited availability \\ \hline
\end{tabular}
\end{table*}

B5G/6G technology is also anticipated to provide secure and reliable communication in crucial infrastructure systems such as transportation systems, power grids, and healthcare systems that demand low latency and high reliability for the seamless transmission of critical data in real time~\cite{javed2022reliable}. Its security features make it well suited for use in these systems, ensuring the protection and transmission of confidential information without any risk of compromise. Moreover, B5G/6G technology will play a crucial role in ensuring security in large networks, in association with the growing demand for IoT and the increasing number of connected devices. Further, it is also expected to establish a robust and secure infrastructure capable of handling the vast amounts of data generated by these devices, including end-to-end encryption, secure device authentication, and robust network security protocols~\cite{khan2022swarm}.

As shown in Fig. \ref{fig:6G}, the role of UAVs in B5G/6G scenarios is of substantial importance, as flying devices will typically be anticipated to densely populate aerial space, enabling a medium network layer between space networks and ground networks \cite{mozaffari2021toward}. The ever-growing and large-scale deployment of UAV technology in a broad range of applications is anticipated to be a primary component of beyond 6G wireless networks in the next couple of decades. The efficacious support of such an enormous employment of UAV technology will require ensuring secure, reliable, resilient, and cost-efficient wireless connectivity.

\subsubsection{Existing Solutions Leveraging B5G/6G, Blockchain, and UAVs}

The new generation of mobile networks is a residence for edge communication technologies that provide unprecedented performance and capabilities~\cite{KUMARI2021102}. Aggarwal \textit{et al.} \cite{aggarwal2020blockchain} considered the evolution of the 6G network and its applications in various areas and the potential of blockchain characteristics in UAV communication. Moreover, the authors proposed solutions and future issues according to the integrations of blockchain-based UAVs and 6G networks. Raja~\textit{et al.}~\cite{raja2022nexus} discussed a nexus of 6G and blockchain for authentication (NBA) system to overcome the loss of authentication data due to node capture and lack of sensor location verification. The NBA system uses permissioned blockchain-based UAVs and intelligent sensor networks to detect code tampering. 

Kalla~\textit{et al.}~\cite{kalla2022emerging} considered four directions of the 6G Internet, namely, Trust-based Secure Networks (TBSN), Harmonized Mobile Networks (HMN), Hyper Intelligent Networks (HIN), and Resource Efficient Networks (REN), through blockchain-based 6G Internet. Kalla~\textit{et al.}~\cite{kalla2022emerging} also discussed that these four categories could characterize future technologies that utilize 6G networks. Additionally, AI/ML-based hyper-intelligent networks focus on artificial intelligence by 6g design. To implement 6G grids, the research identifies significant scientific and technical difficulties and possible rulings for the 6G blockchain. 

Pokhrel \cite{10.1145/3414045.3415949} presented drone-assisted federated learning with blockchain at the 6G edge for the emergency response scenario. Through simulation and modeling, in\cite{10.1145/3414045.3415949}, the author has shown that transmission parameters like power and the number of miners significantly impact the overall energy consumption of a blockchain.  

UAVs are expected to play a critical role in the 6G network, as they are expected to occupy airspace densely and act as a network layer between ground and space networks. Khan \textit{et al.} \cite{khan2022swarm} have discussed security and privacy, AI/ML, and energy efficiency solutions paired with 6G UAV networks. They provided an overview of how a 6G wireless line can integrate blockchain and AI/ML with UAV networks. Further, the review results were presented together with the probable difficulties. Kumar \textit{et al.}~\cite{10.1145/3555661.3560861}
discussed the security aspect of secure UAV communication and designed a novel secure authentication and key agreement framework. The proposed framework preserves the four-factor authentication of an entity, namely unique identity, pseudo-identity, certificate, and timestamp. Additionally, the framework supports establishing session keys with mutual authentication to secure communication between entities.  

Decentralization, transparency, and immutability are the critical missions of blockchain technology that can improve the system's security. Gupta \textit{et al.} \cite{gupta2021blockchain} demonstrated a secure interconnection of UAVs along with blockchain support via the 6G mobile network. García-Magariño \textit{et al.}~\cite{garcia2019security} presented a blockchain-based security mechanism for detecting compromised UAVs in UAV networks. This mechanism also uses a trust policy that relies on information received from most UAVs. Using a secure signature system based on asymmetric encryption and blockchain records, it is impossible to sign false control on behalf of other drones. A compromised drone can enter fake information with its real ID, but the trust policy solves this problem. 

 Shah~\textit{et al.}~\cite{shah2022blockchain} discussed industrial applications of 6G and blockchain technologies, followed by open issues and challenges. Further, Rana \textit{et al.} \cite{8776613} considered blockchain for security purposes to avoid problems related to data leakage or loss during transfer between UAVs. The data transfer process occurs within a blockchain framework that allows all user information to be stored and records to be shared for security management.

\subsection{Deployment Concerns}
\subsubsection{Data amount and speed}
The development of UAVs directly aligns with collecting the data generated with these devices \cite{nex2022uav}. Maturing the environment of UAVs will further pursue the need to develop real-time machine learning (RTML) stability. As machine learning (ML) and deep learning (DL) continue to develop and become more secure, the more capable the outcome desired will be. Nonetheless, there will always be a constant threat of data poisoning, which the infrastructure may not be attainable to defend against. UAV technology is expected to become more dependent on ML algorithms; the justification for secure ML programming and practices is a mandatory indication of how serious the impact a threat actor can possess in the event of a data poisoning attack against UAV systems.

\subsubsection{Network availability}
Wireless network Availability is constructed to relate directly to the decision-making and analysis of AI-supported UAV systems. Verizon issued a statement regarding the use case of 5G and B5G/6G networking communication to become a more viable procurement mechanism that will accompany the advancement of self-driving vehicles and UAVs. With technology as promising as 6G towards UAV environments, the architecture and machine infrastructure are not conditionally ready to be incorporated in mass deployment. The ever-growing deployment of 6G in UAV systems is still far away, as communities must continue undergoing intensive testing of proactive security practices. With these challenges, a larger-functioning backbone for high-frequency data is needed to analyze and adopt 6G into the real-time decision-making capabilities for UAV systems.

\subsubsection{Underdeveloped infrastructure}
The containment and transference of data must be viewed through a more secure approach \cite{arabi2018data}. Notably, the categorical solution of data transference must be sequential to the overall processing that a UAV can produce. The computation of data at the rate of production UAV may not satisfy the desired security. From a convenience standpoint, users and consumers would not be able to tell the difference until the unfortunate event that an accident or tragedy occurs. There is no telling what version of a disaster is imminent when a security model is issued with proper research, simulations, testing, and procurement.


\section{Security Concerns of UAV-based Architectures}
\label{sec:security}

With the rise in the usage of UAVs, various UAV-based architectures have been developed and tailored for various applications. However, the security of these architectures needs to be taken into account to guarantee the safety and confidentiality of transmitted data. This section investigates the security issues with UAV-based architectures, such as potential threats and vulnerabilities. Additionally, we will analyze the strategies and methods utilized to counteract these concerns and explore the challenges associated with securing UAV-based architectures. A better understanding of these security concerns could lead to developing more reliable and secure UAV-based architectures, helping to prevent potential security breaches.

Wang et al.~\cite{wang_improving_2017} developed an iterative algorithm as a solution for Physical Layer Security enhancement using UAV enabled mobile relaying technique. This technique uses the difference-of-concave (DC) program to prove Karush-Kuhn-Tucker (KKT) conditions for guaranteeing convergence, thereby enhancing the secrecy rate. In~\cite{he_communication_2017}, He~\textit{et al.} proposed a communication security mechanism for low-cost implementation for safeguarding against the WiFi and GPS spoofing attack on the targeted UAV. Wang~\textit{et al.}~\cite{wang_energy-efficient_2018} reported their experimental simulation results using an iterative algorithm based on Dinkel-bach's method and sequential convex programming (SCP) for secrecy energy efficiency maximization (SEEM) as a solution to UAV's trajectory planning problem.

Zhu~\textit{et al.}~\cite{zhu_secrecy_2018} used the Matérn hardcore point process in the 3-D antenna gain to improve the secrecy performance of UAV-enabled millimeter-wave networks. Here, the transmit jamming strategy is used to enhance the secrecy performance by considering the practical constraints in deploying UAV communications. To combat the spoofing attack in the UAV control signals~\cite{huang_combating_2018} uses a physical layer approach to determine the source of the control signal packets. In the developed approach, a generalized log-likelihood ratio test framework is constructed to handle the signal spoofing attack of the control signals effectively. According to ~\cite{li_mobile_2018}, there are solutions to maximize the secrecy rate of UAV communication using UAV-aided mobile jamming. Based on their experimentation, a block coordinate descent-based iterative algorithm is used to solve the non-convex optimization problem to achieve better secrecy rate gain.

A dual UAV-enabled secure communication system is introduced in~\cite{cai_dual-uav-enabled_2018}. Moreover, the joint optimization algorithm is developed for managing the jamming caused by multiple UAVs using orthogonal time-division multiple access. From the results, better performance is observed through the usage of joint optimization algorithms. Hierarchical way of intrusion Detection and Response method~\cite{sedjelmaci_hierarchical_2018} provides a high detection rate while monitoring the UAV behavior with lower false positives, lower communication overhead, and increased network density observed through extensive simulations. In contrast, the work in~\cite{lee_uav-aided_2018} creates a UAV-enabled secure communication system for sending private messages to the users at ground stations. This approach using block successive upper bound minimization techniques efficiently solves the non-convex problem, providing better user scheduling and an improved minimum secrecy rate.

Liu~\textit{et al.}~\cite{liu_opportunistic_2018} explained achieving secrecy outage metrics for secure low-altitude UAV swarm communications. This scheme uses eavesdropping probability, backhaul reliability, and Monte Carlo simulations for estimating the secrecy outage probability considering the UAV cooperation, Nakagami-m fading parameters, and eavesdropping probability. Three-dimensional spatial deployment~\cite{zhou_improving_2018} has been introduced to enhance the Physical Layer Security using UAV-friendly jamming based on the outage probability of the intended receiver. In this process, a new security measure is developed based on the eavesdropper's intercept probability (IP), reducing the complexity of an exhaustive search.

To enhance secure transmission in UAV-aided multiple-input-single-output multiple-eavesdropper (MISOME) systems, secure cooperative transmission is presented along with their functionalities~\cite{chen_uav-aided_2019}. In this scheme, an optimal UAV placement strategy is proposed to enhance the secrecy rate and secure energy efficiency performance in the presence of cooperative jamming along with noise beamforming. Even with partial information about eavesdropping, secure UAV communications can be established using cooperative jamming for effective monitoring~\cite{li_cooperative_2019}. The results are observed from the physical-layer security perspective using an iterative optimization method called block coordinate descent, in which the system's worst-case secrecy rate (WCSR) is improved.

Ma~\textit{et al.}~\cite{ma_secure_2019} utilized a cooperative jamming scheme for securing the millimeter-wave communication using UAV enabled relay mechanism. They use multiple UAV-enabled relays and jammers to degrade the signal quality of eavesdropping channels to improve the physical layer security with better secrecy outage probability. Sun~\textit{et al.}\cite{sun_physical_2019} also used an optimal UAV position for maximizing the secrecy rate with the existence of homogeneous Poisson point process (HPPP) eavesdroppers being independent. Consequently, the millimeter-wave transmissions are secured with the support of the derived SWIPT policy for the UAV-based relay systems.

It is likely that for a Secure Amplify and Forward System, physical-layer security and transmission reliability~\cite{tatar_mamaghani_performance_2019} will be necessary for providing desired secrecy metrics for low-altitude UAVs. Li~\textit{et al.}~\cite{li_uav-enabled_2019} designed a novel UAV-enabled mobile jamming scheme using Cooperative Jamming for ground wiretap channel, which provides results with a significant increase in the secrecy rate of the wiretap system.

Further, Zhong~\textit{et al.}~\cite{zhong_secure_2019} overlooked the work done on Secure UAV Communication with a new cooperative jamming approach for defending against eavesdropping. In this technique, two scenarios are considered for experimentation, and using successive convex approximation and alternating optimization techniques, secure communication is ensured. To maintain security in UAV networks, Li~\textit{et al.}~\cite{li_joint_2019} developed an iterative algorithm to solve the non-convex optimization problem. The joint optimization was performed on UAV flight trajectory, and the results provided an improved minimum secrecy rate compared to conventional schemes.

Communication between UAV-to-UAV systems must be secured to avoid intrusion from other random UAVs. In~\cite{ye_secure_2019}, the authors investigated the signal-to-noise ratio (SNR) over the source UAV and the intended UAV receiver through the proposed analytical models with simulations using the Monte-Carlo technique. Based on the idea of UAV Secure Transmission with caching and joint optimization of time scheduling the trajectory of UAV~\cite{cheng_uav-relaying-assisted_2019}, an iterative algorithm is put forward through successive convex optimization that provides a benchmark scheme for maximizing the average secrecy rate.

Based on the analysis of hybrid outage probability for safeguarding UAV Communications,~\cite{liu_safeguarding_2019} proposes a secure transmission scheme. This scheme is employed on the wiretap channel, where the source avoids the eavesdropper and communicates only with the legitimate UAV, providing optimal power allocation despite the full-duplex and active eavesdropper in the network. Secrecy in the presence of UAV Jammer and random UAV Eavesdroppers is imparted using secure connection probability (SCP) and piecewise function described in~\cite{tang_secrecy_2019}. This technique approximates the line-of-sight probability for the communication links and uses sigmoid-based fitting to provide better approximation and secure connection.

The efficiency of the multi-hop transmission can be exploited with an optimal number of hops for secure UAV communication~\cite{wang_secrecy_2020}. In this case, multi-hop network parameters are tuned carefully to optimize the throughput and establish secrecy against UAV surveillance and covert communication. In the work done by~\cite{zhou_secure_2020}, UAV-enabled mobile edge computing using an iterative algorithm of low complexity is designed to optimize the offloading ratio, transmitting power with better secrecy capacity. 


\section{Security Support and Enforcement Aspects} \label{sec:security-support}

The security and privacy services give users the right to control the data and resources accessed by the UAVs in the network. In other words, such security enforcement enables the owners of the data and resources to streamline and disclose their information. This is generally realized by letting the users estimate the energy demands, technological requirements, and security policies. In this section, we investigate the requirements of provisioning robust security support, its importance for the UAVs engaged in communication services, energy demands, power optimization, modern technological support, and state-of-the-art security enforcement algorithms. 

\subsection{Energy and Power Optimization Towards Security Enforcement} \label{subsec:energy}

Security challenges in the physical layer are tackled using UAV mobility through its trajectory design as presented in~\cite{cui_robust_2018}, where a power design methodology for secure UAV Communications is presented. This strategy uses a successive convex optimization method, S-procedure, and the block coordinate descent method for efficiency improvement in the average worst-case secrecy rate. Successive convex approximation (SCA) and alternating optimization (AO) provide a suboptimal solution while trying to optimize the transmit power and trajectory to secure against the UAV over the no-fly zones (NFZs)~\cite{gao_joint_2019}, and the simulation results provide a better computational efficiency resource allocation model and a multi-tenant configuration environment.

Effective power allocation in UAV-enabled relaying systems was implemented by~\cite{sun_power_2019} with slack variables and difference-of-concave (DC) programming. Better secrecy rate maximization and optimized source/relay transmission power are obtained with accurate location information.

By adopting the advantage of coordinated multiple points (CoMP) for strengthening the Physical Layer Security in UAV Swarm~\cite{wang_uav_2019}, which is an aerial network that considers the power allocation strategy using large-scale channel state information (CSI). Based on the observed results, the effectiveness of the proposed scheme is validated in a trajectory-oriented manner. Joint control of trajectory and power is presented in~\cite{zhang_securing_2019}, where the UAV communications are secured using the physical layer security technique. In this joint control mechanism, the average secrecy rates of transmissions are maximized, and the trajectory of the UAV is optimized.

Energy-efficient jamming strategy from~\cite{li_energy_2019}, wherein connectivity of legitimate UAV is established to track the intruding UAVs using tracking algorithms implemented using a new simulation framework. This surveillance outcome ensures better accuracy in tracking suspicious UAVs and an effective packet eavesdropping rate.

Secure communications are enabled in the physical layer of the UAV network by formulating an optimization technique to improve the average secrecy rate as presented and discussed in~\cite{zhou_uav-enabled_2019}. The results are obtained by decomposing the optimization task and solving using successive convex approximation techniques and an alternating iterative algorithm. For UAV mobile-edge computing systems, an energy-efficient computation structure with offloading technique is proposed~\cite{bai_energy-efficient_2019}. The deployment was based on surveillance and traffic management, from which three offloading solutions obtained optimal solutions under different scenarios.

Secure data transmissions in millimeter-wave Simultaneous Wireless Information and Power Transfer (SWIPT) UAV were done based on relay networks~\cite{sun_secure_2019}, and also adopting a combination of design parameters of the multi-hop network in the physical layer to increase the reliability in transmission and reduce the leakage of information with optimal hop counts. Recently, Zhou~\textit{et al.}~\cite{zhou_robust_2020} proposed an optimization technique for securing the UAV-enabled cognitive radio (CR) networks. The developed iterative algorithm is based on S-Procedure, and Bernstein-type inequalities are targeted for optimizing the transmission power and trajectory of the UAV.

\subsection{Modern Technology Support for Security Enhancement } \label{subsec:technology}

As the future UAV network services are anticipated to become ad-hoc and more decentralized, they are primarily vulnerable to cyber-physical attacks and DoS. Recently, AI techniques and other modern supporting technologies have been used as practical solutions to prevent and avoid attacks. In this section, we review the applications of modern supporting technologies in enhancing security challenges. 

\subsubsection{Role of IoT towards security enhancement }
Imparting security features for industrial Internet of Things (IIoT) devices used UAV-enabled friendly jamming scheme was developed by Wang~\textit{et al.}~\cite{wang_uav-enabled_2019} using an anti-eavesdropping scheme on multiple UAVs for analyzing the local as well as overall eavesdropping probability.

One of the solutions for safeguarding the UAV IoT Communications at their core physical layer is based on the implementation presented in~\cite{lei_safeguarding_2020} where the cumulative distribution functions (CDFs) of eavesdropping UAV and main UAV are derived. From the analysis using the signal-to-noise ratio (SNR), the average secrecy rate and the secrecy outage probability are derived to ensure secure UAV IoT Communications.

\begin{itemize}
    \item Machine learning: For establishing cellular connectivity to UAVs, wireless and security challenges need to be mitigated to guarantee secured operation~\cite{challita_machine_2019}, particularly for UAV-based intelligent transportation systems, delivery systems, and multimedia streaming applications. The authors used artificial neural networks and machine learning solutions to guarantee secure real-time communication in UAVs through cellular networks.
    \item Deep learning: Deep neural networks are used for employing strategies on anti-intelligent UAV jamming in~\cite{gao_anti-intelligent_2020}. In this work, the authors developed a deep Q-networks (DQN) capable of operating in a two-dimension space and combining the network with the Markov decision process, which was partially observable. From the results, optimal communication trajectory is obtained with reduced time complexity in UAV jamming applications.
    \item Federated learning: The open nature of 5G technology poses threats against the safe data sharing. Specifically, leakage may produce severe losses for users. Federated learning can provide a promising solution to address some of these challenges. Federated learning allows data models to be shared instead of raw data, thereby reducing the risk of privacy leakage. However, the traditional federated learning frameworks have certain limitations related to the centralized aggregation server, poisoning attacks, and communication barriers. It's worth noting that research and development in the field of federated learning are actively addressing these limitations to improve the privacy, security, and scalability of the framework. As the technology evolves, we can expect advancements that further enhance the safety and usability of federated learning in the context of 5G networks. Notably, Feng et al. \cite{feng2021blockchain} propose a blockchain-empowered decentralized horizontal federated learning framework. This framework enables authentication of cross-domain UAVs by using multi-signature smart contracts. Instead of a centralized server, smart contracts are also used by this framework to compute the global model updates.
\end{itemize}

\subsection{Existing Algorithms Related to UAV Security} \label{subsec:security-overview}

Security is a critical aspect of any digital or communication-enabled system. In UAV-supported communication systems, security is an even more severe challenge. This is mainly due to the system's remote wireless communication and unmanned nature. For instance, unlike terrestrial base stations, if a flying cellular base station is compromised or under an adversarial attack, its serving user equipment may potentially lose cellular connectivity, and the UAV itself is likely to crash \cite{fotouhi2019survey}.

Another security issue is the potential interference in UAV systems, where the line of sight (LOS) links can cause significant interference to cellular user equipment served by the terrestrial base station if an adversary compromises the UAV. Ensuring UAV environment security, especially in cases where drones are deployed for cellular communications, is highly important. This section reviews existing algorithms and solutions to improve and enforce security aspects in UAV systems.

\subsubsection{Routing and trajectory}
Routing is a grand challenge in UAV-aided communication networks due to the dynamic and rapid changes in topology, frequent link disconnections, and high mobility of drones. Thus, designing and implementing reliable routing protocols/solutions that fulfill the minimum overhead and delay constraints face various limitations and challenges, including detecting malicious vehicles.

Designing efficient routing mechanisms for VANETs, which comprise sets of vehicles connected via wireless mediums, is a highly challenging task due to the characteristics and applications of the networks \cite{fatemidokht2021efficient}. Ad hoc mode-oriented routing protocols exist between UAVs and vehicles and between UAVs. The work by Fatemi et al. \cite{fatemidokht2021efficient} presents a type of routing protocol that leverages two different protocols, supporting communication between UAVs and vehicles, as well as between UAVs themselves. The routing approach works by appraising the density of vehicles in a particular road segment based on "Hello" messages exchanged between vehicles. A trust-enabled algorithm is further used collaboratively with UAVs to identify illegitimate vehicles changing pseudonyms.

Despite their advantages in terms of easy deployment and high mobility, UAVs pose a challenge to the security of information associated with air-to-ground line-of-sight communication channels. The broadcast nature of these channels is a security concern that needs to be addressed. Cui et al.~\cite{cui_robust_2018} address this physical layer-level challenge by exploring the mobility of UAVs through trajectory design. In their work, the UAV-ground communication platform is considered with multiple eavesdroppers located on the ground, and the location information of these eavesdroppers is imperfect. An optimization problem is established to maximize the secrecy rate of the platform by enhancing the trajectory planning and power transmission of the UAV within a given flight window.

Moreover, Zhang et al.~\cite{zhang_securing_2019} also address physical-layer security in UAV communication systems by examining both uplink and downlink communications with a ground node, namely ground-to-UAV (G2U) and UAV-to-ground (U2G) communications, which are susceptible to eavesdropping. Instead of relying on ground nodes at fixed or quasi-static locations for security, Zhang et al.~\cite{zhang_securing_2019} use the high mobility of UAVs to proactively establish secure and insecure channels for legitimate and illegitimate links (e.g., eavesdroppers) through trajectory design. Iterative optimization algorithms like successive convex optimization and block coordinate descent can be used to improve the secrecy rates of the G2U and U2G transmissions.

Efficient trajectory and resource allocation are crucial for providing a trustworthy and energy-efficient downlink communication platform for UAVs. In a notable research work by Cai et al.~\cite{cai2020joint}, a joint optimization algorithm is proposed to improve resource allocation, trajectory planning, and jamming policy, ultimately maximizing energy efficiency. To enhance UAV networks and navigation, Krishna et al.~\cite{krishnan2020implementation} proposed an optimization algorithm that dynamically adjusts the trajectory to avoid obstacles and secure the UAV navigation task. In another study, Li et al.~\cite{li2019joint} proposed an algorithm that maximizes the minimum secrecy rate while ensuring fairness among ground terminals (GTs) served by UAVs.

\subsubsection{Resource allocation}
The expectations for future wireless communication technologies, including massive connectivity, low latency, and ultra-high data rates, have dramatically increased. This has led to significant challenges in wireless communication technologies and their associated facilities. Fortunately, UAV-assisted communication platforms offer a promising solution \cite{fotouhi2019survey, zeng2019accessing} that can alleviate the limitations of conventional wireless communications at the physical layer. By leveraging the high mobility and flexibility of UAVs, the performance of the communication platform can be improved by moving the UAVs closer to end users. However, despite this potential, technical challenges still need to be addressed to realize the promised performance gains fully.

Rigorous power limitations primarily hinder effective UAV communications. Due to UAVs' size and weight limitations, their onboard batteries' energy storage capacity is typically small. Furthermore, the power consumption of flight and communication systems heavily depends on the UAV's velocity and trajectory. Therefore, enhancing the energy efficiency of UAVs has garnered tremendous interest and research efforts in the literature. Notably, Cai et al. \cite{cai2020joint} propose a joint trajectory and resource allocation approach to improve UAV communication systems' energy consumption and security. The proposed approach optimizes the trajectory of the information UAV, resource allocation, and jammer UAV's jamming policy to maximize the overall system energy efficiency.

Li et al. \cite{li2020resource} propose a resource allocation scheme for secure multi-UAV communication systems that relies on multi-purpose base stations dispatched to ensure secure communications with authorized ground users in the presence of eavesdroppers. The authors leverage orthogonal frequency division multiple access (OFDMA) to enable active base stations to communicate with legitimate ground users through assigned subcarriers. By jointly optimizing the UAV's trajectory and communication subcarrier allocation policy, the solution ensures fairness in secure communication and maximizes the minimum secrecy per user.

Hong et al. \cite{hong2019resource} proposed a method for resource allocation in a secure UAV-assisted simultaneous wireless information and power transfer (SWIPT) system in the presence of multiple eavesdroppers. They aim to maximize the per-user secrecy rate by jointly optimizing the trajectory and transmission power of the UAV while solving the non-convex problem. Xu et al. \cite{xu2020joint} developed a method to maximize the minimum secure computing capacity for both non-orthogonal multiple access (NOMA) and time division multiple access (TDMA) in UAV-assisted mobile edge computing (MEC) systems. They achieve this by jointly optimizing the UAV trajectory, communication resources, and computation resources. These works highlight the importance of joint optimization of UAV trajectory, transmission power, and resource allocation for achieving efficient and secure communication in UAV networks. Finally, another research interest in this area is resource allocation for multi-UAV collaborative tasks, which can be approached as a combinatorial optimization problem and solved using an improved clustering methodology. The work by Fu et al. \cite{fu2019secure} addresses this optimization problem while allowing multiple UAVs to work collaboratively and complete tasks with lower energy consumption.

\subsubsection{Authentication and encryption}
UAVs are now used in various fields, such as rescue and search missions, military operations, and monitoring tasks. However, security remains a significant concern, as UAVs handle sensitive data related to individuals' privacy that adversaries should not compromise or disclose. Researchers have addressed specific security vulnerabilities in UAV systems, including Yoon et al.'s study \cite{yoon2017security}, which proposed a solution to prevent hijacking attempts against commercial UAVs' network channels. The authors presented an algorithm to maintain control of UAVs in hijacking situations and establish a security-enhanced communication channel with an authentication layer.

To ensure the reliable, resilient, rapid, and secure delivery of services to end-users through Drones as a Service (DaaS), rigorous guidelines are required for designing a 5G-UAV network. One possible solution is to leverage blockchain consensus and its decentralization feature, where UAVs can act as mobile access points, service suppliers, or routing nodes. Aloqaily et al. \cite{aloqaily2021design} propose a 5G network-supported environment based on UAVs and blockchain to fulfill fluctuating user demands with access supply. The proposed solution facilitates decentralized service delivery (DaaS) and the capability of routing to/from end-users in a trustworthy way. A remarkable feature of this work is that it ensures authenticated information communication, which achieves a high delivery success rate.

Furthermore, despite the widespread use of IoD in various applications, ranging from civilian to military, privacy and security remain critical challenges that must be adequately addressed during communication, whether with the moving access points in the sky or the control station server. One solution to this problem in IoD is presented by Bera et al. \cite{bera2020blockchain}, a blockchain-enabled secure data management scheme that provides resistance against various threats in IoD communication entities.

The data link performance is another critical metric in UAVs that could be potentially affected by the communication channel, which could ultimately impact the overall reliability of data transmission. Furthermore, wireless communication channels can also be a surface for various remote attacks \cite{atoev2019secure}. Therefore, it is essential to ensure that the wireless communication channels between the ground control station (GCS) and vehicles are resilient while maintaining data link performance efficiency by employing lightweight security authentication solutions. For instance, the lightweight security authentication mechanism for UAV environments \cite{teng2019lightweight} and secure UAV systems in the presence of eavesdropping UAV nodes \cite{zhou2019secure} are viable solutions to ensure secure communication channels.

Artificial intelligence (AI) can be integrated with blockchain technology to enhance the underlying networking infrastructure of UAVs, including 5G communications. Notably, Gupta et al. \cite{gupta2021fusion} present an AI-based architecture design to secure 5G communications in the drone environment. Specifically, the presented design leverages InterPlanetary File System (IPFS) as a platform for information storage, enhancing overall communication privacy and security, underlying network performance, and reducing transaction storage overhead while facilitating on-the-fly decision-making. Wazid et al. \cite{wazid2020private} propose another solution incorporating AI techniques and private blockchain infrastructure to secure communications in IoT-based drone-aided healthcare services. AI was also leveraged by Zhang et al. \cite{zhang2020uav} to address the trajectory design challenge for multi-UAV cooperation scenarios through a multi-agent reinforcement learning approach.

\begin{table*}[h!]
\caption{A taxonomy of existing solutions for UAV-enabled secure communication architectures.} 
\label{tab:security-works}
\centering
\begin{tabular}{|p{1.0cm}|p{3.5cm}|p{5.0cm}|p{5.0cm}|}
\hline
\textbf{Reference} 
& \textbf{Solution}
& \textbf{Key contributions}
& \textbf{Implications}\\ \hline
\hline
~\cite{faraji2021secure} &  Smart agents &  Establishes secure communication between UAVs with the support of smart agents  & Implements a three-step negotiation process to identify reliable neighbors, and discard the traffic generated by the malicious UAVs   \\ \hline
~\cite{li2018protecting} & Q-learning-based algorithm &  Increase the system secrecy capacity, and decrease the attack rate  &  Smartly handles the attacks even under imperfect estimation of the channel with the support of Nash equilibrium \\ \hline
~\cite{khan2022secure} & Secure communication protocol & Leverage vulnerabilities in Mavlink packets by encrypting through additional security layers & The compromised security requirements were overcome through Mavlink as one of the lightweight open-source protocols \\ \hline
~\cite{khan2020emerging} & Secure communication protocol for UAVs &  Overview of UAVs, their communication protocols, and security threats &  The proposed secure communication protocol can enhance the security of UAVs and encourage their wider adoption \\ \hline
~\cite{jain2021enabling} & AI-based UAV-borne secure communication framework for Industry 5.0  &  The proposed model involves a novel image steganography technique with optimal pixel selection and encryption processes for classification of securely received UAV images  &  The presented model shows good performance indicating the potential for enhancing the security and reliability of UAV communication in Industry 5.0 and other safety-critical applications \\ \hline
~\cite{fotohi2020agent} & SID-UAV method to secure UAV-to-UAV communications and resist malicious UAVs &  Self-matching system that detects the safest path between UAVs & The proposed method outperforms other existing methods indicating its effectiveness in enhancing the security of UAV networks \\ \hline
~\cite{huang2020energy} & 3D trajectory optimization model for a solar-powered UAV  &  Secure communication against eavesdroppers while avoiding no-fly zones and considering energy consumption and solar power harvesting  & Rapidly-exploring Random Tree (RRT) method is developed to construct the UAV trajectory, and computer simulations show that the proposed method outperforms a baseline method \\ \hline
~\cite{rashid2019secure} & Security model based on Identity Based authentication scheme for UAV-integrated HetNets & A scheme that it is resistant to vulnerabilities of intruders such as replay and impersonation  &  The proposed scheme can help to mitigate security issues in UAV-integrated HetNets, particularly in military scenarios, and ensure secure communication among military users using the network \\ \hline
\end{tabular}
\end{table*}


\section{Blockchain-based Security Architecture for UAV in B5G/6G} \label{sec:blockchain}

This section summarizes the typical blockchain-based security architectures meant for UAVs in conjugation with the B5G/6G services. The presented solutions are based on existing studies and reflect UAV systems' fundamental ideas and mechanisms. However, it is unfortunate that only some of the horizons of the related proposed architectures in the literature could be covered due to the unique intention of each framework in terms of the design, navigating the environment, and wireless communication strategies.

As of today, due to the heterogeneous network infrastructure and integrated ubiquitous devices in B5G/6G services, the security demands of B5G/6G networks are to be integrated with blockchain services. Given this, Nguyen~\textit{et al.}, \cite{nguyen2019blockchain} surveyed the potential of utilizing blockchain to protect the 5G services from external attacks. Anticipating the B5G/6G demands and securing its services, Praveen et al.~\cite{praveen2020blockchain} identified the specific areas where blockchain could enhance the privacy and security of 5G services. Mehta~\textit{et al.}~\cite{mehta2020blockchain} summarize the taxonomy of security challenges in 5G-enabled UAVs and provide security solutions using blockchain for industrial applications. Further, the authors have also presented a model for Multi-Operator network slicing and implemented the same, securing the 5G services using blockchain frameworks. 

The survey in~\cite{ullah2020cognition} presented the regulations and standardization of UAVs for potential applications and the energy harvesting, collision avoidance, channel modeling, and security issues that can ensure new business openings. Further, to enhance the security within UAV networks, Alladi et al.~\cite{alladi2020applications} reported using blockchain. In this study, the authors addressed the key enabling technologies and various applications of blockchain for deployment in UAV networks. In particular, the challenges existing in UAV-based surveillance, network security, inventory management, and decentralized storage were considered for analysis with blockchain integration and suggested a few research directions. 

\subsection{Spectrum Management}
As the development of blockchain-based technology is getting popular, they are also used by UAV operators for secure spectrum trading~\cite{qiu_blockchain-based_2020}. Here, an incentive-based spectrum blockchain framework is proposed for improving the efficiency of the spectrum trading environment for UAV-assisted cellular networks. Distribution of underutilized spectrum to human-to-human users and machine-to-machine (M2M) are performed using 5G heterogeneous networks considering the privacy ensured using blockchain framework\cite{zhou2020blockchain}. Moreover, this work addresses the efficiency of the spectrum allocation, along with the secure sharing of spectrum and incentive mechanism with a relevant case study. 

Qiu et al. \cite{qiu2019blockchain} were among the first to pioneer blockchain-enabled spectrum trading for UAVs. The researchers offered the solution from the mobile network operator perspective, which could be used for security assessment in spectrum trading in UAV-enabled cellular networks. The authors observed that blockchain technology significantly improves the security aspects of spectrum sharing.

\begin{figure}[!t]
\centering
\centerline{\includegraphics[width=0.5\textwidth]{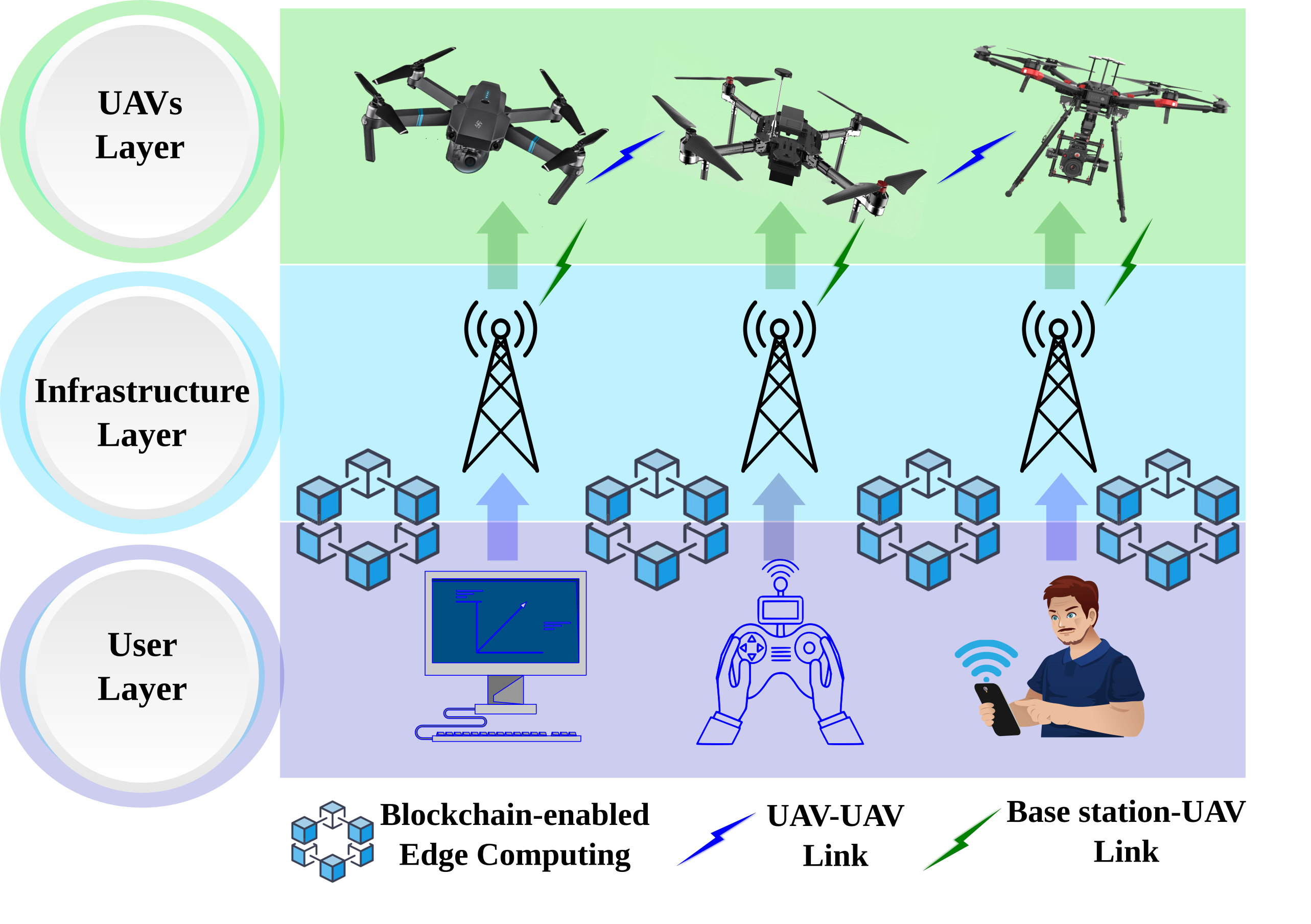}}
\caption{Blockchain-enabled edge computing for UAVs in the junction of user and infrastructure layers.}
\label{fig:networking}
\end{figure}

\subsection{Secured Path Planning}
Specific UAVs demand marked paths in their environment, few others require cleared flight paths, and some need to be effectively managed in a 3D space. Considering all such constraints, Dasu et al. \cite{dasu2018geofences} implemented geofencing and established air-traffic control based on different modes by herding the UAVs with the support of blockchains and 5G service. 

\subsection{Privacy Preservation and Intrusion Detection}
From the intrusion detection perspective, the blockchain-based delivery system through UAV services, named DeliveryCoin~\cite{ferrag2019deliverycoin} employs short signatures and hash functions. The authors achieved a consensus inside the delivery platform through the UAV-driven forwarding schemes by applying this privacy-preserving scheme. The proposed intrusion detection systems in each 5G cell could handle network attacks and improve the effectiveness of the UAV communication services. 

In~\cite{chamola2020comprehensive}, Chamola et al. reviewed the impact of 5G services and blockchain technology for assessing the health implication due to the outbreak of COVID-19. They have highlighted its effects on the global economy in association with other supporting technologies such as IoT, UAVs, and AI towards mitigating the COVID-19 outbreak. Inspired by the work mentioned above, Islam and Shin \cite{islam2020blockchain} developed a blockchain-based secure healthcare scheme with the support of UAVs and IoT. Here, the UAVs communicate with body sensor hives via tokens and share the keys, enabling low-power and secure communication through 5G services. 

\subsection{Cooperative Navigation Strategies}
Cooperated blockchain-5G networks are summarized with the open-end research challenges persisting in this domain. A similar work reported by Li~\textit{et al.}~\cite{li2018uav} surveyed the usage of 5G services based on UAV platforms with the joint consideration of operation in various layers and communication strategies. The game theoretical approach of blockchain-based spectrum sharing, introduced by Choi et al.~\cite{choi2019game}, uses a tit-for-tat strategy to establish user cooperation in the 5G-Enabled IoT network. The strategy's performance was assessed in a dense network by encouraging the users to share the information regularly. Through this technique, it was evident that the spectrum-sharing features are optimized compared to the traditional approaches.
 
\subsection{Secured IoT Services for UAVs}
Operating UAVs in aerial mode and base station mode demands a confidential exchange of messages. So ensuring the best security mechanisms in the physical layer enhances the secured and controlled mobility with better performance~\cite{li2019secure}. Each UAV in the network of UAVs acts as a Blockchain node that holds onboard functionality for observing and interpreting block transactions and can exchange information with other UAV nodes in the network~\cite{kuzmin2018blockchain}. Blockchain technology is also widely used to address the challenges in IoT services, mainly driven through 5G services. Dai~\textit{et al.}~\cite{dai2019blockchain} introduced the term blockchain of things (BCoT) and presented its architecture. Further, the authors elaborated on the implications of this technology along with 5G services targeted for industrial applications. 

Nguyen~\textit{et al.}\cite{nguyen2020blockchain} surveyed cooperated blockchain-5G networks by compiling the works in this domain related to IoT, transportation, healthcare, and UAVs. Similarly, 5G-enabled IoT in industrial applications with blockchain technology as its backbone is surveyed in~\cite{mistry2020blockchain}, where Mistry~\textit{et al.} shed more light on the privacy preservation of autonomous vehicles towards secured automation. Subsequently, Ge~\textit{et al.}~\cite{ge2020semi} propose a novel IoT communication architecture based on blockchain technology. Their solution implements a distributed scheme with blockchain's privacy and security benefits while mitigating computation and storage requirements. 

\subsection{Secure Data Management}
A blockchain-empowered security feature is enabled in a UAV swarm for data acquisition tasks from IoT devices and further routed to the closest node in the UAV swarm \cite{islam_bus_2019}. Here, the sender is validated using the digital signature algorithm, and the UAV swarm implements the $\pi$-hash bloom filter with two-phase validation. For the blockchain-based, IoT, and 5G-enabled UAV environment, a secure data delivery and collection scheme was proposed by~\cite{bera2020blockchain}, aiming to secure data management among the UAV communication. As the blockchain-based secure framework needs to manage several potential attacks to impart better security and functionalities, the computation and communication overheads need considerable attention. The proposed architecture is based on decentralized nature that performs the security analysis with considerably moderate communication efforts and less computational overheads than conventional schemes. Secure data acquisition scheme from IoT devices enabled using blockchain concept was used in a UAV swarm \cite{islam_bus_2019}. In this scheme, the IoT devices forward the encrypted data to the UAV swarm, providing secure connectivity and better energy consumption.
 
\subsection{Edge Computing}
Edge computing, along with 5G and UAV, may provide an opportunity for new business values for the telecom sector. According to \cite{sharma2019neural}, ultra-reliable communication is employed in mobile edge computing by utilizing efficient cache-enabled UAVs as on-demand nodes. Here, the deployed neural-blockchain-based provides a flat architecture with reliable means of communication. The work in \cite{miatton2020blockchain} proposed using distributed ledger technology with 5G for unlocking the new values of utilizing UAVs for connected supply chains. However, research needs to be carried out to ensure the validity of the blockchain involved in edge computing services on data storage, retrieval, and processing. Guan et al.~\cite{guan2020blockchain} proposed their blockchain-based distributed security solution for the UAV-enabled mobile edge computing model that reduces fraud occurrences and reasonably rewards authenticated participants. The framework comprises a smart contract, and the sub-blockchain enhances the system's stability. Further, based on the conceptualized framework, the results show that the proposed scheme consumes minimum resources and time for computation.

\subsection{Trust Protocol for Internet of Drones}
Mershad et al.~\cite{mershad2022proact} propose a framework that assumes the existence of multiple UAV networks that can connect to each other via ground control stations (GCSs). A central or distributed authority controls each group of GCSs. The blockchain network consists of UAVs and GCSs cooperating to generate and store blockchain blocks. The proposed system assumes that certain types of data will be stored within blockchain blocks saved at drones and GCSs, while other categories of data will be saved within blocks that only need to be stored at the GCSs. Two main types of transactions exist in the blockchain: the first type will contain complete data related to the transaction. In contrast, the second type will include transaction metadata and a reference to the actual data stored on an external storage system.


\section{Open Issues, Challenges and Future Trends in the Security of UAV Architecture} \label{sec:issues}

The section provides a summary of key findings and lessons learned from previous research in the field of UAV security. It also outlines some of the open issues and challenges that still need to be addressed, such as developing secure communication protocols and implementing robust authentication mechanisms. Furthermore, it may discuss future trends and potential solutions for addressing these challenges, such as using machine learning algorithms for anomaly detection and developing advanced encryption techniques.

\subsection{Summary of Key Findings and Lessons Learned} \label{subsec:summary}

The ongoing research in enhancing security in UAV-based systems involves applying iterative algorithms, joint optimization techniques, and cooperative jamming. These methods improve secrecy rate, reduce communication overhead, and increase network density while providing desired secrecy metrics for low-altitude UAVs. In addition, the development of dual UAV-enabled secure communication systems and hierarchical intrusion detection and response methods has also demonstrated improved performance.

Due to UAVs' unmanned and remote wireless nature, ensuring security in UAV-based communication systems is critical. Compromised UAVs can result in loss of connectivity and interference. In this article, we discuss different essential solutions for solving this problem. The solutions presented include routing with trust-enabled algorithms, mobility, and optimization, physical layer security via trusted and distrusted channels, secrecy rates, and resource allocation for energy efficiency and secure navigation. However, challenges still exist, such as fairness and limitations of the physical layer. 

Blockchain and 5G services have shown great potential in addressing these challenges, including secure spectrum trading, path planning, privacy preservation, and intrusion detection systems for UAVs. They are also utilized for cooperation in 5G-enabled IoT networks and secure data management in UAV swarms, among other applications. By using UAVs as blockchain nodes and employing a game-theoretical approach for efficient spectrum sharing, blockchain can serve as a backbone for secure and private data management while mitigating computation and storage requirements in IoT communication architecture.

\subsection{Open Issues and Challenges} \label{subsec:challenges}
Despite the existence of many articles in the literature, including this one, covering and categorizing a broad range of solutions to the different identified security issues and threats, it is anticipated that we will see more specifically designed UAV solutions. The expected security designs will pay more attention to the low node density (e.g., flying ad-hoc networks), restricted power consumption, highly dynamic topology changes, highly dynamic mobility of nodes, and high dependence on radio propagation.

 Although we categorized many security solutions in this article specifically for UAV network architectures, many of these solutions need to be more suitable and flexible for UAVs with high mobility, i.e., high-speed movement. Furthermore, UAVs' standardized communication channels for authorization, authentication, and access control must be improved in flying ad-hoc networks.

\subsubsection{Security and Privacy} 
Apart from enhancing security, it is crucial to prioritize protecting sensitive information, safeguarding against potential malicious attacks, and establishing robust measures to prevent unauthorized access to the blockchain network. Ensuring privacy is a critical aspect that necessitates carefully considering and implementing encryption techniques and access controls. Moreover, the design of a secure UAV architecture must account for the dynamic nature of UAVs, which are susceptible to physical tampering and signal interference. Therefore, incorporating mechanisms for tamper-proofing the UAVs and implementing secure communication protocols are essential elements to mitigate potential vulnerabilities. Additionally, maintaining the integrity and immutability of data stored within the blockchain is paramount, necessitating the establishment of stringent auditing and verification mechanisms. By addressing these challenges, a secure UAV architecture utilizing blockchain technology can foster trust, reliability, and enhanced security in UAV communication, leading to expanded applications and advancements in next-generation wireless networks.

Furthermore, UAVs generate and exchange sensitive data, including flight telemetry, sensor data, and mission-related information. Ensuring the security and privacy of this data is paramount. Integrating blockchain technology can provide tamper-resistant and transparent data storage, but challenges remain in designing robust security mechanisms, authentication protocols, and encryption techniques tailored for UAVs in the context of blockchain-5G integration.

\subsubsection{Regulatory and Legal Considerations}
The utilization of UAVs in wireless networks operates within the framework of regulations and legal standards. Integrating blockchain technology into this context introduces a new set of considerations, including additional compliance requirements and challenges pertaining to data ownership, liability, and accountability. Adhering to regulatory frameworks is vital to ensure UAVs' responsible and lawful operation in wireless networks. Blockchain-based systems introduce unique data ownership challenges, as multiple entities may contribute to and access the blockchain network. As a result, clear protocols and agreements must be established to define data ownership and the corresponding rights and responsibilities.

Moreover, liability concerns may arise when using UAVs equipped with blockchain technology, particularly in scenarios involving data breaches or accidents. Determining the appropriate parties responsible for such incidents and allocating liability can be complex. Therefore, a comprehensive legal framework that considers the implications of blockchain integration within UAV operations is necessary. Accountability is another key aspect that needs to be addressed. Establishing mechanisms to trace and attribute actions performed within the blockchain network is crucial for maintaining transparency and ensuring accountability among the involved parties. By navigating the complexities of regulations, data ownership, liability, and accountability, integrating blockchain technology into UAVs in wireless networks can promote responsible and compliant use, fostering trust among stakeholders and facilitating the advancement of this technology in various industries.

Last, UAVs are subject to various regulations and legal requirements, particularly in terms of flight operations, privacy, and data handling. Integrating blockchain and 5G introduces additional considerations for compliance and governance. Addressing these legal and regulatory challenges, ensuring compliance with existing frameworks, and developing new guidelines specific to blockchain-5G integrated UAV systems is essential.

\subsubsection{Standardization}
Standardization plays a crucial role in developing secure UAV architectures incorporating blockchain technology. Establishing standardized protocols and frameworks is imperative to ensure seamless interoperability, compatibility, and widespread adoption across diverse platforms and vendors. By implementing standardized practices, the UAV industry can overcome barriers such as fragmentation and proprietary solutions, facilitating a more cohesive ecosystem. Standardization efforts should focus on defining standard communication protocols, data formats, and security measures that enable UAVs and blockchain systems to interact efficiently and securely. Moreover, standardized frameworks promote collaboration and knowledge-sharing among stakeholders, enabling the development of innovative solutions and fostering a healthy competitive environment. These standards also facilitate regulatory compliance, as regulatory bodies can refer to established protocols and frameworks when assessing the security and reliability of UAV architectures. Furthermore, standardization promotes scalability and future-proofing by ensuring that UAV architectures incorporating blockchain technology evolve and integrate seamlessly with emerging technologies and advancements.

\subsubsection{Trust and Governance}
Building trust and implementing effective governance models are vital to ensuring the secure operation of UAVs within a blockchain network. Establishing trust among the participating nodes is essential, fostering a reliable and collaborative environment. This can be achieved by developing consensus mechanisms that enable nodes to collectively agree on the validity of transactions and ensure the integrity of the blockchain data. By addressing issues related to node reputation, such as evaluating the reliability and performance of individual nodes, the overall trustworthiness of the UAV architecture can be enhanced. Trust management mechanisms, including reputation systems and identity verification, play a crucial role in establishing trust among network participants. Robust and transparent governance models are also necessary to govern the interactions and decision-making processes within the blockchain network. Defining clear rules and procedures for participation, consensus, and dispute resolution ensures a fair and accountable environment for all stakeholders. Moreover, governance frameworks need to consider the evolving nature of UAV technology and adapt accordingly to accommodate emerging challenges and opportunities. By emphasizing trust and governance, a secure UAV architecture utilizing blockchain technology can inspire confidence, foster collaboration, and enable UAVs' safe and reliable operation in a wide range of applications and industries.

\subsubsection{Scalability and Interoperability} 
Blockchain networks, such as the popular public blockchain Ethereum, face scalability issues when it comes to handling a large number of transactions. UAVs generate vast amounts of data that need to be securely stored and processed in real-time. Developing scalable blockchain solutions capable of handling the high transaction volume and data throughput from UAVs is an ongoing challenge.

Moreover, UAVs are often part of larger systems, such as IoT networks or smart city infrastructures. Ensuring interoperability between different blockchain platforms, 5G/6G networks, and other technologies involved in UAV operations is crucial for seamless integration. Developing standards and protocols that allow for interoperability and data exchange between different systems will be a significant challenge.

\subsubsection{Latency and Real-Time Communication} 
UAVs require low-latency and reliable communication for tasks like remote control, navigation, and data transmission. While 5G networks offer high bandwidth and reduced latency, ensuring seamless and uninterrupted communication between UAVs and blockchain networks is still a challenge. Optimizing protocols and network architectures to support real-time communication and synchronization with blockchain systems is crucial.

\subsubsection{Energy Efficiency} 
UAVs operate on limited battery power, and energy efficiency is a critical factor for their endurance and overall performance. Both blockchain and 5G technologies consume significant energy resources. Finding ways to optimize the energy consumption of blockchain systems and 5G networks in the context of UAV applications can enhance their practicality and extend flight times.

Overall, the integration of blockchain technology and 5G/6G networks into UAVs presents remarkable opportunities, but it also requires addressing various technical, security, regulatory, and interoperability challenges. Continued research, collaboration between different stakeholders, and iterative development are essential to overcome these obstacles and unlock the full potential of this integration.


\subsection{Recent Trends and Future Directions} \label{subsec:trends}
In this subsection, we highlight the future research directions in accordance with the current open challenges and issues pertaining to UAV networks and how blockchain technology can help alleviate some of these open challenges. Although the BC-aided B5G/6G UAV architecture enhances the different UAV services by empowering privacy, security, trust, and improving resource management, integrating such an architecture faces several challenges, detailed next.

\subsubsection{QoS and scalability}
The existing literature contains many research works demonstrating blockchain technology's throughput, scalability, and delay/latency. This is mainly due to the increasing volume of data communicated through the network, which is continuously growing. Additionally, the consensus mechanisms in blockchain can affect the overall Quality of Service (QoS) of the system in terms of throughput, scalability, and delay, as the rate of data generation is usually faster than the mining process. Scalability has always been a major concern in blockchain technology, and researchers must continue to seek improvements to enhance the QoS of blockchain-assisted systems, such as UAVs.

\subsubsection{Federated learning and UAV}
5G technology provides an excellent communication infrastructure to support a wide range of intelligent services and applications in UAVs. For example, drones flying near end-users can act as relay nodes to transmit data and strengthen edge servers. Moreover, through federated learning (FL) techniques, UAVs can assist in processing and managing collected data and sharing learned models with cloud/fog nodes for further aggregation and comprehensive comparison to aid decision-making. However, using this approach often incurs additional computation overhead and energy consumption, which requires optimization of UAV resources (such as the battery, CPU, node election, and formation) based on advanced task scheduling and allocation.

\subsubsection{Diversity of data}
In UAV systems, devices (i.e., drones) typically exchange a broad range of message types, which must be handled differently. Some of these messages are used to instruct drones to complete their tasks, such as IoT data created by the UAV, data exchanged among UAVs, instruction data received from the pilot, and other data generated from other UAV system entities and relayed by flying drones. Therefore, future blockchain-assisted UAV systems should consider the diversity of UAV data by adopting more flexible transactional systems while preserving the security and privacy of the various data types.

\subsubsection{Considerations for privacy \& security in the design phase}
When designing B5G/6G UAV architectures with blockchain assistance or developing blockchain-enabled applications for deployment in UAV networks such as the IoD, privacy, and security should be considered essential components. Incorporating privacy and security approaches is highly recommended as best practice to avoid or reduce threats. Furthermore, forensic mechanisms can be adopted to design blockchain-assisted UAV systems. Forensic-by-design, for instance, is a promising mechanism that can be incorporated since it has been proven effective in other cyber-physical environments and networks. This forensic mechanism is particularly suitable for tracing attack sources and reconstructing threat events.

\subsubsection{Real-time isolation}
Although localization is critical when direct UAV location measurement is unavailable, localization errors can persist. Therefore, future blockchain-assisted 5G UAV systems need an approach that efficiently and quickly isolates the compromised device exposed to security and privacy breaches. This approach will serve as a fault isolation solution, providing a protection layer against a holistic failure of the UAV network. Furthermore, since the line of sight connection can be lost at any moment, future UAV systems are expected to provide a strategy that facilitates the drone's return to the initial base station upon the occurrence of this situation.

\subsubsection{Collision and obstacle avoidance}
Despite the significant advances in UAV technology to address drone collisions and obstacle avoidance, the potential for crashing drones remains one of the most significant risks associated with this technology for both properties and people, and drones may even collide with flying planes. As a result, it is crucial for the existing IoD to carefully consider obstacle detection and deploy real-time, reliable solutions to handle the presence of obstacles. If blockchain technology is deployed for UAV 5G infrastructures, it is essential to have a comprehensive understanding of the range of IoD communication services.

\subsubsection{Computer vision}
At present, on-board surrounding analytical systems are being widely used, leading to a significant increase in energy consumption and weight for the devices. If blockchain technology is deployed in UAV systems, advanced and sophisticated computer vision tools must be integrated to stream the surrounding views, such as images, and feed them back to the object recognition entity for feature processing and analysis. Moreover, leveraging computer vision tools with blockchain technology can potentially increase the UAV systems' computational complexity and resource requirements, making it critical to optimize the system for energy consumption and storage capacity. Therefore, researchers must develop efficient algorithms and techniques to improve the computational efficiency of computer vision tools in blockchain-assisted UAV systems. Additionally, optimizing UAV devices' power management and storage capacity will be crucial to support the increased computational demand while ensuring the system's longevity.

\subsubsection{AI-enabled blockchain attack mitigation}
The IoD network is now facing several new and sophisticated attacks, which require implementing artificial intelligence (AI) based mitigating mechanisms. AI techniques such as artificial neural networks, deep learning, and machine learning algorithms can be used to optimize the security and privacy of the IoD network. However, implementing these techniques presents significant challenges, and selecting the appropriate AI algorithm for specific security and privacy requirements still requires further research and exploration. In addition, it is crucial to monitor the IoD network for emerging threats continuously and to adapt the AI-based mitigating mechanisms to address new attack vectors. This requires a proactive approach to security and privacy that can quickly respond to changing circumstances and protect the network from evolving threats.

\subsubsection{Trustworthy data aggregation}
As drones collect and incorporate a large amount of data in the network, securing and encapsulating this data within a single unit is essential before sending it. Future blockchain-assisted UAV 5G architectures can deploy flexible and secure aggregation mechanisms for drone data. Aggregating this data before sending it can reduce communication and energy costs, which are critical for UAV network resources. Cryptographic techniques can be leveraged to aggregate multiple ciphertexts into a single unit, simultaneously enforcing access control and confidentiality of the aggregated data. Such techniques can be used to create a secure and efficient mechanism for drone data aggregation.

\subsubsection{Intrusion detection}
Integrating blockchain technology into a UAV 5G system may introduce additional security threats to its network, which are specific to the blockchain. Therefore, intelligent and robust intrusion detection and prevention systems (IDPSs) are required for such a network. UAV researchers are recommended to take advantage of the recent advances in intrusion detection solutions to design and develop more robust and more intelligent IDPSs to mitigate existing threats in UAV systems and the threats that may arise from integrating blockchain technology. It is crucial to ensure the security and privacy of UAV systems, as they are becoming increasingly prevalent in various fields and industries. Moreover, developing effective security measures for UAV 5G blockchain systems is essential not only to protect against potential cyber attacks but also to establish trust and confidence among users of these systems.

\subsubsection{Standardization and regulatory policies}
Ensuring standardization and regulatory compliance remains a significant concern for both UAVs and blockchain technologies. To regulate and standardize the usage and integration of UAV devices, particularly drones, in urban areas, future policies and laws must be designed and enforced for privacy, safety, and security purposes. Additionally, regulatory rules are necessary to control service providers' and entities' access to specific data in the decentralized ledger. Efficient Implementation of these policies and regulations requires collaboration among industry stakeholders, policymakers, and regulatory bodies. Close cooperation can help ensure that UAVs and blockchain technologies are used and integrated in a manner that promotes public safety and respects privacy while also advancing the development of these emerging technologies.

\subsubsection{Off-chain storage}
As mentioned earlier, UAV technology facilitates the exchange of various data types or messages. Some of this data may be significant in size, making it difficult to store in the ledger, especially if it requires frequent deletion or modification. To address this challenge and improve the overall performance of the UAV system, off-chain storage may be a viable solution. Off-chain storage also provides the added benefit of reducing the blockchain's computational workload and network traffic, which can enhance the system's speed and efficiency. Additionally, off-chain storage can help minimize the costs associated with storing large amounts of data on the blockchain, making it a more cost-effective solution in certain scenarios.

\subsubsection{Location and divisibility}
Ensuring mutual agreement or consensus on allocating services and location among service providers is another significant challenge for UAV technology. Failure to address this challenge could impact the operability of UAV technology and hinder its large-scale deployment. Furthermore, it could delay the integration of future broadband cellular networks, including 6G, requiring seamless coordination and collaboration among various stakeholders. 

To overcome this challenge, it is necessary to establish clear guidelines and regulations for allocating services and locations among service providers. These guidelines should take into account the various requirements and constraints of different stakeholders, including UAV operators, service providers, and regulators. By fostering greater collaboration and cooperation among these stakeholders, it is possible to enhance the operability and deployment of UAV technology while ensuring that it remains compatible with future broadband cellular networks.

\section{Conclusion} \label{sec:conclusion}

This review article offers a comprehensive overview of UAV-based B5G/6G communication technologies and architectures that support seamless services. While UAVs were once considered a hobbyist's tool, they now play a critical role in disaster management and emergency response and can even provide wireless communication services. As the demand for establishing seamless B5G/6G-based communication services to remote areas increases, drones and UAV technology are becoming more popular. However, it is crucial to address security concerns in the research and development of advanced UAV-based 5G architecture. To address this need, this article conducts a comprehensive review of the security aspects of UAVs in the context of B5G/6G system architecture, including enabling technologies, privacy issues, and security integration at all protocol stack layers. Additionally, this survey analyzes existing mechanisms to secure UAV-based B5G/6G communications and energy and power optimization factors. The article also summarizes the use of modern technological trends for establishing security and protecting UAV-based systems, along with identifying open challenges and strategies for future research in this area.

\printunsrtglossaries

\end{document}